\documentclass[twoside,twocolumn,english,prc, twocolumn, showpacs]{revtex4}
\usepackage[T1]{fontenc}
\usepackage[latin1]{inputenc}
\usepackage{amsmath}
\usepackage{graphicx}
\usepackage{amssymb}

\makeatletter
\@ifundefined{textcolor}{}
{%
 \definecolor{BLACK}{gray}{0}
 \definecolor{WHITE}{gray}{1}
 \definecolor{RED}{rgb}{1,0,0}
 \definecolor{GREEN}{rgb}{0,1,0}
 \definecolor{BLUE}{rgb}{0,0,1}
 \definecolor{CYAN}{cmyk}{1,0,0,0}
 \definecolor{MAGENTA}{cmyk}{0,1,0,0}
 \definecolor{YELLOW}{cmyk}{0,0,1,0}
 }

\makeatletter

\usepackage{amsfonts}

\makeatother

\makeatother

\makeatother

\usepackage{babel}

\begin{document}

\title{Quantum Tunneling and Scattering of a Composite Object:\\
Revisited and Reassessed}

\author{Naureen Ahsan}
\email{na05@fsu.edu}

\author{Alexander Volya }
\email{avolya@fsu.edu}
\homepage{www.volya.net}

\affiliation{Department of Physics, Florida State University, Tallahassee, Florida
32306}
\begin{abstract}
This work presents an extensive exploration of scattering and tunneling
involving composite objects with intrinsic degrees of freedom. We
aim at exact solutions to such scattering problems. Along this path
we demonstrate solution to model Hamiltonians, and develop different
techniques for addressing these complex reaction-physics problems,
discuss their applicability, and investigate the relevant convergence
issues. As examples, we study the scattering of a two-constituent
deuteron-like systems either with an infinite set of intrinsic bound
states or with a continuum of states that allows for breakup. We show
that the internal degrees of freedom of the projectile and its virtual
excitation in the course of reactions play an important role in shaping
the S-matrix and related observables, giving rise to enhanced or reduced
tunneling in various situations.
\end{abstract}


\maketitle

\section{Introduction}

Reaction physics involving composite objects is a major and critical
subject while encountered in the context of processes like fusion,
fission, particle decay, as well as specific branches of science including
chemistry, atomic physics, condensed matter physics, etc. In all these
phenomena, more often than not, the scattering or tunneling object
has its own degrees of freedom. Various pertinent scenarios have been
explored earlier, where the tunneling has been shown to be enhanced
by the additional degree(s) of freedom, which may have arisen from
the compositeness of the object \cite{Flambaum:2005,Bertulani:2007,Balantekin:1998,Bonini:1999,Goodvin:2005,Kayanuma:1994,Bacca:2006},
from its interaction with another particle \cite{Ivlev:2004}, or
directly from quantum field excitations \cite{Flambaum:1999}.

This is a complicated and generally non-perturbative problem, involving
vastly different scales. While there are many techniques and methods
of dealing with this problem, most of them involve simplifications.
For example, some studies of the models that are similar to ours involve
restriction on the range of energy of the projectile \cite{Bertulani:2007},
the mass-ratio of the constituents \cite{Bertulani:2007,Goodvin:2005,Kayanuma:1994,Bacca:2006},
the number of states available in the intrinsic system \cite{Goodvin:2005,Bacca:2006},
etc. In addition, most models exclude the possibility of virtual excitations
of the object undergoing a reaction \cite{Flambaum:2005,Goodvin:2005,Bacca:2006}.
While simplifications work well at times, it is also common that the
{}``slightly simplified'' problem turns out to be very different
from the original one. Moreover, some formally exact techniques, as
demonstrated in Ref. \cite{Ahsan:2010} and further discussed in
this paper, do not necessarily provide a path to a convergent solution
for an arbitrary subset of parameters. The paramount goal of this
work is to find an exact solution to a given reaction problem, which
is free from the above mentioned limitations and is reliably convergent.

In order to reach this goal we limit our studies specifically to a
problem in one dimension and to a composite object with two constituents,
only one of which interacts with an external potential. A deuteron
hitting a Coulomb barrier could be a fair example of such a projectile.
This picture has been modeled in several different ways in our work.
However, the techniques that we develop and the study of how they
work are general and are not limited, in their applicability, to our
examples or models only. Moreover, we believe that many of our findings
are generic and there are realistic situations that can be represented
by even these simple models \cite{Lemasson:2009}. 

Our discussion is organized as follows. In Sec. \ref{sec:General}
we start by identifying our models and invoking some definitions of
reaction physics. Then in Sec. \ref{sec:The-Projection-Method} we
examine a particularly simple example of a deuteron-like system reflecting
from an infinite wall. This case provides an excellent illustration
of the pivotal role of virtual excitations in the dynamics. It also
shows how the formally exact method of projecting the reaction dynamics
onto the intrinsic shell-model-like space could fail to yield reliable
results. We put forward and demonstrate the Variable Phase Method
(VPM) in Sec. \ref{sec:The VPM}, followed by solutions to various
examples in Sec. \ref{sec:Applications}. While the VPM has been used
by others in the past, we extend it so as to include virtual channels.
This novel extension requires us to explore the role of virtual channels
and to discuss the convergence of solutions with the number of virtual
excitations included in consideration. This is done in Sec. \ref{sec:Role-of-virtual}.
A study of scattering and breakup of a system with a continuum of
states is presented in Sec. \ref{sec:continuum}. The summary and
conclusions are laid out in Sec. \ref{sec:Summary-and-conclusions}.

\section{General Description of the Problem \label{sec:General}}

\subsection{The model}

Throughout this text we examine a one-dimensional problem. We consider
a projectile which is a composite object made up of two particles
which have masses $m_{1}$ and $m_{2}$ and are bound by an intrinsic
potential $v(x_{1}-x_{2})$, where the particle coordinates are $x_{1}$
and $x_{2}$, respectively. This composite system interacts with an
external potential $V(x_{1},x_{2})$. The usual center-of-mass and
relative coordinates are\begin{equation}
X=\frac{m_{1}x_{1}+m_{2}x_{2}}{M}\,,\qquad x=x_{1}-x_{2},\label{eq:CM_X}\end{equation}
and the corresponding total and reduced masses are
\begin{equation}
M=m_{1}+m_{2},\qquad m=\frac{m_{1}m_{2}}{m_{1}+m_{2}}.\label{eq:CM_P}\end{equation}
The Hamiltonian for the system can be written as 
\begin{equation}
H=-\frac{\hbar^{2}}{2M}\frac{\partial^{2}}{\partial X^{2}}+V(x_{1},x_{2})+h\label{eq:Hamiltonian}\end{equation}
 where the intrinsic Hamiltonian \begin{equation}
h=-\frac{\hbar^{2}}{2m}\frac{\partial^{2}}{\partial x^{2}}+v(x)\label{eq:intrinsicH}\end{equation}
has eigenstates $\psi_{n}(x)$ with the corresponding energy eigenvalues
$\varepsilon_{n}$: \begin{equation}
h\psi_{n}(x)=\varepsilon_{n}\psi_{n}(x),\qquad n=0,1,2,...\label{eq:inHsolution}\end{equation}

The general scattering problem can be formulated in a traditional
way, using the asymptotic form of the wave function.
At $X\rightarrow\mp\infty$ it is given by
\begin{subequations}
\begin{equation}
\Phi(X,x)  \simeq \frac{e^{iK_{n'}X}}{\sqrt{|K_{n'}|}}\psi_{n'}(x)+\underset{n\in\mathrm{open}}{\sum}\frac{R_{nn'}}{\sqrt{|K_{n}|}}e^{-iK_{n}X}\psi_{n}(x)\,,\label{eq:asympt-}
\end{equation}
\begin{equation}
\Phi(X,x) \simeq \underset{n\in\mathrm{open}}{\sum}\frac{T_{nn'}}{\sqrt{|K_{n}|}}e^{iK_{n}X}\psi_{n}(x)\label{eq:asympt+}\end{equation}
\label{eq:asympt}
\end{subequations}
respectively. The wave function above corresponds to an
incident beam coming from the left with a particle in the intrinsic
state (channel) $n'.$ Here $K_{n}$ is the center-of-mass momentum
of the system with total energy $E_{T}$ while in channel $n$, \begin{equation}
K_{n}(E_{T})=\frac{1}{\hbar}\sqrt{2M(E_{T}-\varepsilon_{n})}.\label{eq:define_K}\end{equation}
Here, and later in this paper, the symbol `$\simeq$' is used to indicate
an asymptotic equality that involves only open channels $n$ with
$E_{T}>\varepsilon_{n}$. Contributions from the closed channels,
for which $E_{T}<\varepsilon_{n}$, decay exponentially with distance
from the scattering potential and are not present in the asymptotic
form. The coefficients $R$ and $T$ are referred to as reflection
and transmission amplitudes due to their physical meanings. $|R_{nn'}|^{2}$
and $|T_{nn'}|^{2}$ represent the probabilities for the incoming
beam in channel $n'$ to reflect and transmit, respectively, in channel
$n.$ The conservation of probability hence implies

\begin{equation}
\underset{n\in\mathrm{open}}{\sum}\left(|R_{nn'}|^{2}+|T_{nn'}|^{2}\right)=1.\label{eq:unitarity}\end{equation}
It should be mentioned that for the scattering problem to
be fully determined one should consider, in addition to \eqref{eq:asympt},
an incident beam coming from the right, which gives rise to another
set of reflection and transmission amplitudes. If and when it is necessary
to distinguish between these two, the amplitudes in Eqs. \eqref{eq:asympt}
for the incident beam coming from the left and traveling in the positive
$x$ direction are denoted by $R_{+}$ and $T_{+}$ instead of just
$R$ and $T$; $R_{-}$ and $T_{-}$ are used when they are associated
with an incident beam traveling from the right to the left.

\subsection{The $S$-matrix\label{sub:Smatrix}}

While it is convenient to use the reflection and transmission amplitudes,
the formal $S$-matrix is still essential for establishing a relation
between this description and the traditional scattering theory. In
addition, $S$-matrix allows one to utilize the symmetries of the
problem, and determine relations among the amplitudes. Despite $S$-matrix
being a textbook subject \cite{Merzbacher:1998,Lipkin:1973}, there
are a few non-trivial features that emerge in the case of coupled-channel
problems and with non-symmetric potentials \cite{Nogami:1996,Kiers:1996,van_Dijk:2008}.
We review some of them in what follows.

Let us first consider the case of one open channel, which is of particular
importance for many examples considered in this work. For a real potential
barrier the transmission amplitude is symmetric between the incoming
beam traveling from the left and that from the right, which follows
directly from the complex-conjugated Schrödinger equation, showing
time-reversal invariance. The $S$-matrix can be defined in several
different ways \cite{Nogami:1996}. It is quite common to select
a basis with incoming and outgoing waves so that $S=\mathbf{1}$ at
high energies or in the absence of a potential barrier. An alternative
approach is to choose the $S$-matrix to be symmetric, which is possible
because of the time-reversal invariance. Unfortunately, it is impossible
to accommodate both properties simultaneously. We choose the second
alternative and define the $S$-matrix using the following symmetric
and asymmetric (in space) asymptotic forms of the incoming waves at
$|X|\rightarrow\infty$, \[
\Phi^{+}(X)\simeq\frac{i}{\sqrt{2}}\exp\left(-iK|X|\right) \,\,{\rm and}\]
\[
\Phi^{-}(x)\simeq\frac{i}{\sqrt{2}}\frac{X}{|X|}\exp\left(-iK|X|\right).\]
The outgoing-wave basis comprises the corresponding complex-conjugated
wave functions. Then the $S$-matrix in terms of reflection and transmission
amplitudes is\[
S=-\frac{1}{2}\left(\begin{array}{cc}
(R_{+}+R_{-})+2T & (R_{-}-R_{+})\\
(R_{-}-R_{+}) & (R_{+}+R_{-})-2T\end{array}\right).\]
Note that the S-matrix is symmetric, and $T_{+}=T_{-}=T$
due to time-reversal invariance. From unitarity of the $S$-matrix
we find that $|R_{-}|=|R_{+}|,$ $|R_{\pm}^{2}|+|T_{\pm}^{2}|=1,$
and $\Re\left[T^{*}(R_{+}+R_{-})\right]=0$. The convenience with
the above definition is that, for a symmetric potential, $R_{-}=R_{+}$.
Also, parity is a good quantum number, and hence the $S$-matrix is
diagonal with matrix elements
\begin{equation}
S^{\pm}=-(R\pm T)=\exp(2i\delta^{\pm}).\label{eq:define_phases}\end{equation}

An extension of the above discussion to a more general multichannel
case is straight-forward \cite{Kiers:1996,Baz:1971}. From unitarity
it follows that

\begin{equation}
R_{\pm}^{\dagger}R_{\pm}+T_{\pm}^{\dagger}T_{\pm}={\bf 1},\quad R_{\pm}^{\dagger}T_{\mp}+T_{\pm}^{\dagger}R_{\mp}={\bf 0}.\label{eq:unitRT}\end{equation}
Time-reversal invariance leads to \[
T_{\pm}=T_{\mp}^{T},\, R_{\pm}=R_{\pm}^{T}.\]
Finally, reflection symmetry of the scattering potential
leads to $R_{+}={\cal P}R_{-}{\cal P}$ and $T_{+}={\cal P}T_{-}{\cal P}$.
The equalities for one channel are modified due to the different parities
of the intrinsic states of the composite object. ${\cal P}$ denotes
the parity operator in the channel space, so that ${\cal P}^{2}={\bf 1}$
with ${\cal P}_{nn'}=\delta_{nn'}\pi_{n}$, where $\pi_{n}$ is the
parity of the intrinsic state $\psi_{n}(x)$.

In the presence of reflection symmetry it is sufficient to consider
only beams originating from the left and thus to deal only with $R_{+}$
and $T_{+}$. From this point onward we omit the subscript $+$ (hence
returning to the notations used in Eqs. \eqref{eq:asympt}.
The symmetries discussed for the multichannel case are summarized
as follows: \begin{equation}
R=R^{T},\quad T={\cal P}T^{T}{\cal P},\label{eq:parityRT}\end{equation}

\begin{equation}
R^{\dagger}R+T^{\dagger}T={\bf 1},\quad R^{\dagger}{\cal (P}T)+(PT)^{\dagger}R={\bf 0}.\label{eq:relationsRT}\end{equation}
We define \begin{equation}
S^{\pm}=-(R\pm{\cal P}T)\label{eq:S-matrix}\end{equation}
in this case, so that the $S$-matrix is symmetric, and the phase
shifts approach zero in the limit of zero energy, since $R=-1$ and
$T=0$ in this limit. Conditions at other thresholds are related to
Levinson's theorem which, for one-dimensional scattering, is discussed
in Ref. \cite{Kiers:1996}.

\section{The Projection Method: Examples and limitations \label{sec:The-Projection-Method}}

Before we actually describe the Projection Method, let us emphasize
one important issue. While the observed picture (the $S$-matrix for
example) is seen through the asymptotic forms of the wave functions
in the open channels, the crucial dynamics of a scattering process
takes place in the vicinity of the scatterer, and involves virtual
(or closed) channels just as much as the open channels. The virtual
channels are populated in accordance with the time-energy uncertainty,
and lead to an immensely complicated process. Excluding the virtual
channels from consideration could therefore lead to erroneous results
in the observed quantities. A {}``simple'' model example discussed
below elucidates both the complexity and the importance of virtual
excitations. This model constitutes an infinite wall as a scatterer
which interacts with only one of the two constituents of the composite
object. We refer to this model as the \textquotedbl{}deuteron and
Coulomb-wall\textquotedbl{} model, as defined in Sec. \ref{sub:deuteron_wall}.

It is noteworthy that numerous authors \cite{MORO:2000,Sakharuk:1999,Ahsan:2007},
have worked on this subject, but reports of the findings are scarce.
Mathematical issues, difficulties with stability of the solutions,
and lack of appreciation from the scientific audience are some of
the possible reasons.

In the following method, referred to as the Projection Method, a solution
is attempted by projecting the reaction dynamics onto the intrinsic
basis set. In some way this approach is similar to various projection
techniques used in nuclear many-body studies that involve reactions
\cite{Okolowicz:2003,Volya:2006,Volya:2009}.

We start our presentation by returning to the \textquotedbl{}deuteron
and Coulomb-wall'' model and to the Projection Method. We draw some
conclusions regarding the earlier discussions \cite{MORO:2000,Horoi:1999,Sakharuk:1999,Sakharuk:2000,Zelevinsky:2000,Zelevinsky:2005,Ahsan:2007}
by presenting an exact solution, showing limitations of the Projection
Method, and highlighting the overall importance of this example for
the understanding of reaction dynamics and development of techniques. 

\label{truncation}

\subsection{The deuteron and Coulomb-wall model \label{sub:deuteron_wall}}

For all the models in this work we assume that in the composite projectile,
loosely referred to as the deuteron, only the second particle interacts
with the potential, $V(x_{1},x_{2})\rightarrow V(x_{2})$. In this
section we concentrate on an example where the potential is represented
by an infinite wall or, the {}``wall'':
\[
V(x_{2})=\begin{cases}
\infty & \mathrm{\mbox{when}\;}0<x_{2}\\
0 & \mathrm{\mbox{otherwise}}\end{cases}\,.\]
 The traditional textbook methods prescribe looking for a
full wave function in the form \begin{equation}
\Phi(X,x)=\frac{e^{iK_{n'}X}}{\sqrt{|K_{n'}|}}\psi_{n'}(x)+\sum_{n}\frac{R_{nn'}}{\sqrt{|K_{n}|}}e^{-iK_{n}X}\psi_{n}(x),\label{eq:wallWF}\end{equation}
with the boundary condition $\Phi(X,x)=0$ at $x_{2}=0$.
In contrast to the asymptotic form in \eqref{eq:asympt},
where the summation includes open channels only, the sum here is over
all channels and the expression holds for all values of $x_{2}<0$.
The meaning of the reflection amplitudes $R_{nn'}$ is therefore extended
to include the virtual channels as well as open. The asymptotic form
\eqref{eq:asympt} is recovered at $|X|\rightarrow\infty$
because the term corresponding to each virtual channel, say $n$,
decays with distance (from the wall) through the exponential factor
$e^{-|K_{n}X|}$, and therefore does not appear in the asymptotic
sum. This exponent can be expressed in a generic way as $e^{iK_{n}|X|}$
by assuming the principal branch of the square root in Eq. \eqref{eq:define_K}.
The branch being specified allows one to consider momentum in a complex
plane.

The location $x_{2}=0$ in the boundary condition translates into
$x=x_{1}$ and $X=\mu_{1}x$, as follow from Eqs. (\ref{eq:CM_X}).
Here we define relative masses as
\begin{equation}
\mu_{1,2}=m_{1,2}/M,\quad\mathrm{and}\quad\mu=m/M.\label{eq:mass-ratios}\end{equation}
Thus, the equation to be solved is 
\begin{equation}
\Phi(\mu_{1}x,\, x)=0\label{eq:zerosum}\end{equation}
for all $x$'s.

The length scale for this problem is determined by a quantity
$\lambda$ that is associated with the characteristic width of the
intrinsic potential $v(x)$. The intrinsic system also defines the energy scale,
based on the usual coordinate-momentum uncertainty, as \begin{equation}
\epsilon=\frac{\hbar^{2}}{2m\lambda^{2}}.\label{eq:energyunit}\end{equation}
In what follows we use $\lambda$ and $\epsilon$ as our units of
length and energy respectively. This is equivalent to using dimensionless
energy units rescaled to $\epsilon$, namely, $\varepsilon_{n}\rightarrow\varepsilon_{n}/\epsilon$
for the intrinsic energies, and ${E}\rightarrow E/\epsilon$ for the
center-of-mass kinetic energy ; and lengths rescaled to $\lambda$,
namely, ${x}\rightarrow x/\lambda$, $X\to X/\lambda$ and ${K}_{n}\rightarrow K_{n}\lambda$
for coordinate and momentum variables. Thus, it is assumed that $\lambda=1$
and $\epsilon=1$ unless otherwise stated. The center-of-mass kinetic
energy for an incident beam in channel $n$ is $E=E_{T}-\varepsilon_{n}$;
in almost all our examples the incident beam is in the ground state
channel, and therefore $n=0$. 

Truncating the number of channels at some large $N$ and looking for
a solution in the space spanned by the functions $\psi_{n}$ for $n<N$
constitutes the projection approach. Thence emerge the equations \begin{equation}
\sum_{n}\frac{D_{ln}\left[-i\mu_{1}({K}_{n'}+{K}_{n})\right]}{\sqrt{|{K}_{n}|}}\, R_{nn'}=-\frac{\delta_{ln'}}{\sqrt{|{K}_{n'}|}}\,,\label{eq:D-matrix-eqn}\end{equation}
where matrix $D$ is defined as \begin{equation}
D_{ln}(\varkappa)=\int\psi_{l}^{*}(x)\: e^{\varkappa{x}}\:\psi_{n}(x)\, dx.\label{eq:shift-matrix}\end{equation}
This is the expectation value of the momentum shift operator in the
intrinsic basis. Eq. \eqref{eq:D-matrix-eqn} is obtained by projecting
the boundary condition onto the intrinsic basis set. Note that for
a virtual channel, the argument of the $D$-matrix becomes real and
positive.

At beam energies below the first threshold, when only the ground state
channel ($n=0$) is open, Eq. \eqref{eq:zerosum} is particularly
simple, since scattering is characterized by only a single phase of
the reflection amplitude. Due to unitarity $|R_{00}|=1$; and the
single $S$-matrix phase $\delta$ is defined through $e^{2i\delta}=-R_{00}$.
Equation \eqref{eq:zerosum} then reads\begin{equation}
\psi_{0}({x})\,\sin\left[\mu_{1}{K}_{0}{x}-\delta\right]+\frac{1}{2}\:\sum_{n\in\mathrm{closed}}\, R'_{n0}\sqrt{\frac{|{K}_{0}|}{|{K}_{n}|}}e^{\mu_{1}|{K}_{n}|\,{x}}\psi_{n}({x})=0,\label{eq:realzerosum}\end{equation}
where $R'_{n0}=R{}_{n0}e^{-i(\delta+\pi/2)}$ is real for any $n$.

To further illustrate the situation let us review two specific examples
of intrinsic potential $v(x)$ where the eigenstates \eqref{eq:inHsolution}
and the shift matrices \eqref{eq:shift-matrix} can be found analytically.

\subsubsection{Infinite square well (\textquotedbl{}well'') confinement \label{sub:Infinite-square-well}}

In the first example, which is that of an infinitely-deep square well
intrinsic confinement, the length scale $\lambda$ is defined so as
to set the width of the well to $\pi\lambda$, thus 
\begin{equation}
v(x)=\begin{cases}
0 & \mathrm{\mbox{when}\;}|x|<\pi\lambda/2\\
\infty & \mathrm{\mbox{otherwise}}\end{cases}\,.\label{eq:SQWellPot}\end{equation}
The eigenstates and the corresponding energies for this
square well are \begin{equation}
\psi_{n-1}({x})=\sqrt{\frac{2}{\pi}}\,\sin\left[\left({x}+\frac{\pi}{2}\right)n\right],\quad\varepsilon_{n-1}=n^{2}\,,\label{eq:eigenSQWellPot}\end{equation}
where $n=1,\:2,\dots$, so that the indices for both $\psi$ and $\varepsilon$
start from $0$, and the energy scale \eqref{eq:energyunit} is $\epsilon=\varepsilon_{0}$.
The corresponding shift matrix is \begin{equation}
D_{nn'}(\varkappa)=\frac{4nn'\varkappa\left[(-1)^{n+n'}\exp\left(\frac{\pi\varkappa}{2}\right)-\exp\left(-\frac{\pi\varkappa}{2}\right)\right]}{\pi\left[(n+n')^{2}+\varkappa^{2}\right]\left[(n-n')^{2}+\varkappa^{2}\right]}.\label{eq:D_sqwell}\end{equation}

\subsubsection{Harmonic oscillator (\textquotedbl{}HO\textquotedbl{}) confinement
\label{sub:Harmonic-oscillator-confinement}}

One could criticize the infinite square well potential as being too
sharp and therefore leading to nonphysically high intrinsic excitations.
Therefore, the harmonic oscillator intrinsic confinement $v(x)=m\omega^{2}x{}^{2}/2$
is presented as a second example, which does not have this controversial
feature.

The unit of length here is defined by the standard oscillator length,
$\lambda=\sqrt{\hbar/m\omega}$. The eigenstates and the eigenvalues
are defined in terms of the usual Hermite polynomials $H_{n}$,
\begin{equation}
\psi_{n}({x})=\frac{1}{\sqrt{2^{n}n!\sqrt{\pi}}}H_{n}\left({x}\right)\exp\left(-\frac{{x}^{2}}{2}\right)\,,\quad\varepsilon_{n}=(2n+1),\label{eq:sho_wv}\end{equation}
where $n=0,1,\dots$, and the energy unit is $\epsilon=\varepsilon_{0}=\frac{1}{2}\hbar\omega$.
The corresponding shift matrix is 
\begin{equation}
D_{nn'}(\varkappa)=\sqrt{\frac{n_{<}!}{n_{>}!}}\,\left(\frac{\varkappa}{\sqrt{2}}\right)^{|n-n'|}\,\, L_{n_{<}}^{|n-n'|}\left(-\frac{\varkappa^{2}}{2}\right)\,\exp\left(\frac{\varkappa^{2}}{4}\right)\,,\label{eq:D_oscillator}\end{equation}
where the Associated Laguerre Polynomials $L_{n}^{l}$ appear,
with $n_{<}$ and $n_{>}$ denoting the smaller and larger, respectively,
of the two indices $n$ and $n'$.

\subsection{Solutions, difficulties, and limitations \label{sub:Drawbacks}}

From the explicit forms of the shift matrices in the two cases described
in Eqs. \eqref{eq:D_sqwell} and \eqref{eq:D_oscillator}, it is clear
that the shift matrix in Eq. \eqref{eq:D-matrix-eqn}, when inverted,
has highly singular elements for $N\rightarrow\infty$. To be precise,
$\varkappa_{n}\sim n\sqrt{m_{1}/m_{2}}$ for a square well (and $\varkappa_{n}\sim\sqrt{2nm_{1}/m_{2}}$
for oscillator) which implies that the elements in Eq. \eqref{eq:shift-matrix}
of the shift-matrix have exponentially different scales.

\begin{figure}[!ht]
\begin{centering}
\includegraphics[width=7cm]{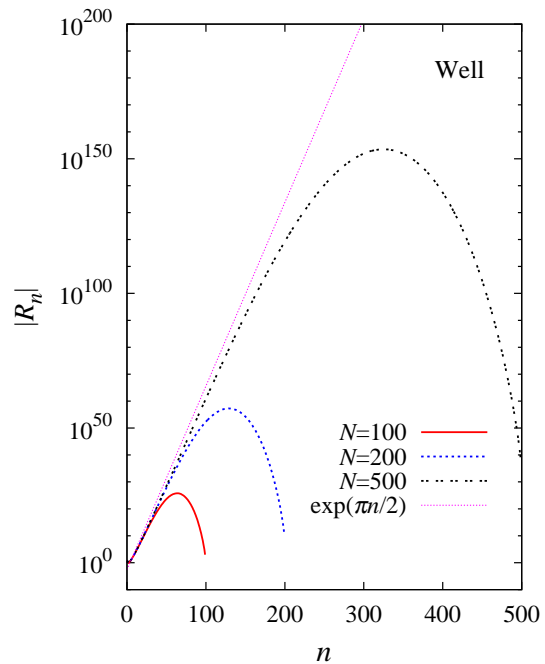} 
\par\end{centering}

\caption{(Color online) \textbf{Well and wall.} Absolute values of reflection
amplitudes for virtual channels are plotted against channel number
$n$. The infinite square well potential (intrinsic) is used in this
example, with $m_{1}=m_{2}$. The incident beam is in the ground state
with kinetic energy $E=0.1$. Different curves correspond to different
truncations $N$. The straight line shows the curve $\exp(\pi n/2)$.
The actual results closely follow this line initially and then deviate
due to truncation in the channel space.\label{virtuals}}

\end{figure}

This difficulty of matrix inversion can be handled by performing a
linear transformation from the set of basis states $\psi_{n}(x)$
to a different set. Transformation to configuration localized state
is discussed in Ref. \cite{MORO:2000}. In our studies we used the
singular value decomposition which is also effective. As was observed
in Refs. \cite{Horoi:2000,Sakharuk:1999,Sakharuk:2000,Zelevinsky:2000,Zelevinsky:2005},
there are complications in numerical convergence with an increased
number $N$ of included channels. The core of the problem is that
the amplitudes for real and virtual channels involve very different
scales. We find that for the square well, for example, remote virtual
channels scale approximately as $R_{n}=R_{n0}\sim\exp\left[\frac{\pi n}{2}\sqrt{\frac{m_{1}}{m_{2}}}\right]$
(similar scaling follows for the oscillator). The behavior of the
virtual coefficients is illustrated in Fig. \ref{virtuals}. Here
and in what follows, the second index in $R_{nn'}$ and $T_{nn'}$,
which corresponds to the incident channel, is dropped when it is the
ground state channel, i.e., when $n'=0$.

\begin{figure}[!ht]
\begin{centering}
\includegraphics[width=8cm]{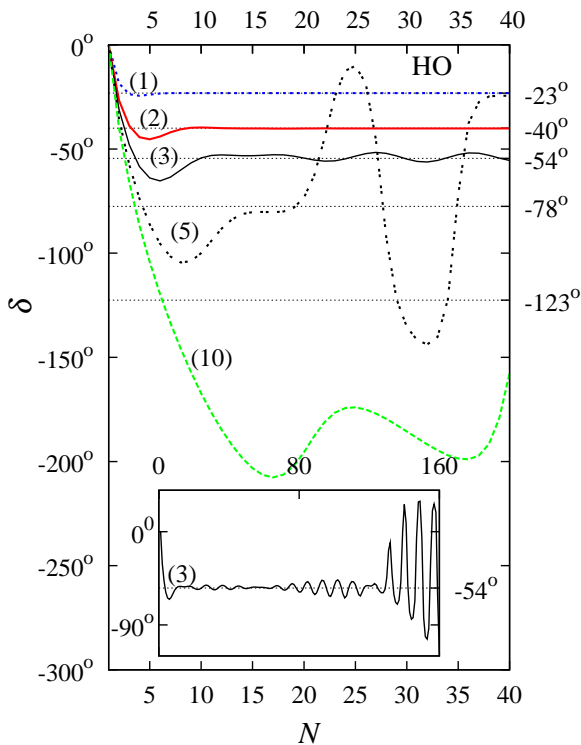} 
\par\end{centering}

\caption{(Color online) \textbf{HO and wall.} This plot demonstrates the failure
of the Projection Method, which uses Eq. \eqref{eq:D-matrix-eqn},
to solve the scattering problem where the system of two particles
bound by a harmonic oscillator confinement collides with an infinite
wall. The incident kinetic energy is $E=1=0.5\,\hbar\omega$, which
means that the total energy is half way between those of the ground
state and of the first excited state. Different curves are labeled
on the graph with the corresponding mass-ratios, $m_{1}/m_{2}=$ 1,
2, 3, 5, and 10. Phase shift $\delta$ is shown to vary with $N$,
the number of channels included in the calculation. The horizontal
grid lines along with the tic-marks on the right indicate the values
of the phase shifts, as obtained in a convergent way with a different
method, the VPM (see Sec. \ref{sec:The VPM}). The curve for $m_{1}/m_{2}=3$
on a large scale up to $N=160$ is shown in the inset. \label{fig:osc_e05_low} }

\end{figure}

As the physics of interest is comprised of small contributions from
exponentially large excitations, the problem has mathematical issues.
This is apparent also from Eqs. \eqref{eq:zerosum} and \eqref{eq:realzerosum},
where one is attempting to make a series with exponentially divergent
coefficients vanish. This condition fails to fulfill especially for
large mass-ratios $m_{1}/m_{2}$, when the coordinate-range $|x|\sim\lambda$
(where $\psi_{n}$ are not zero) implies large exponential factors
$e^{|K_{n}|\lambda\mu_{1}}$. The physics behind this is that when
the interacting particle is stopped by the wall, the non-interacting
component continues its motion until its entire kinetic energy is
converted into virtual intrinsic excitations of the confining potential
which is necessary for the system to be reflected. The bigger the
mass of the non-interacting particle $m_{1}$ relative to $m_{2}$,
the more kinetic energy it has, and the more complicated do the virtual
excitations become. Figure \ref{fig:osc_e05_low} shows how this issue
effects calculated results. 

For {}``good'' mass-ratios, which is roughly when $m_{1}/m_{2}\le2$,
reliable solutions can be obtained \cite{MORO:2000,Ahsan:2007} that
agree with exact and stable solutions gotten through a different method
(the VPM, see Sec. \ref{sec:The VPM}), as shown in Sec. \ref{sub:Infinite-wall}.
For these satisfactory results, the Projection Method had to involve
arbitrary-precision numerics ensuring that both the small and the
large contributions are properly taken care of. The reflection probabilities
in different open channels calculated for the square well and harmonic
oscillator models with $m_{1}/m_{2}=1$ are reliable. They are also
identical to those obtained through the VPM, and are shown in Figs.
\ref{fig:wall-well} and \ref{fig:wall-oscillator}, respectively,
in Sec. \ref{sub:Infinite-wall}. These two figures display cusps
at thresholds, which is a consequence of unitarity \cite{Landau:1981,Baz:1971}.
In addition to that, there are weak oscillations which, as discussed
in the same section, become more pronounced in the case of a more
massive non-interacting particle.

The solution, however, is still elusive. This is especially visible
for {}``bad'' mass-ratios with large values of $m_{1}/m_{2}$. The
Projection Method results are shown in Fig. \ref{fig:osc_e05_low}.
As the mass of the non-interacting particle $m_{1}$ gets larger,
the results become extremely unstable. While a satisfactory value
may be obtained for some cases, the approach is still flawed since
the inclusion of more channels (which must be accompanied by increased
numerical precision) does not necessarily improve the results and
may eventually lead to increasing oscillatory instabilities. This
is demonstrated in the inset where the curve (3) for $m_{1}/m_{2}=3$,
which seems to converge initially, is continued up to N=160 where
its behavior becomes erratic.

In order to solve this problem, one should depart from projection
onto the basis states. In what follows this is achieved by introducing
a Variable Reflection Amplitude $U_{nn'}(X)$ (see Ref. \cite{Razavy:2003})
through the following equation,

\begin{equation}
R_{nn'}=e^{i(K_{n}+K_{n'})X}\left[2i\sqrt{K_{n}K_{n'}}\, U_{nn'}(X)-\delta_{nn'}\right].\label{eq:wall-U}\end{equation}
Then Eq. \eqref{eq:zerosum} takes the form \[
\sum_{n}U_{nn'}(\mu_{1}x)\psi_{n}(x)=0,\]
and can be efficiently solved by selecting a discrete set of $N$
coordinate locations. This is the essence of a different approach
discussed next.

\section{The Variable Phase Method (VPM)\label{sec:The VPM}}

It follows from the discussion in the previous section that the approach
based on projection onto the intrinsic basis is unpredictable in its
ability to handle the problem. As an alternative, the time-dependent
methods have previously been used to treat similar problems \cite{van_Dijk:2008}.
Here we discuss the Variable Phase Method (VPM), which is a well established
technique for treating multi-channel tunneling and scattering. It
dates back to works presented in Refs. \cite{Morse:1933,Drukarev:1949,Kynch:1952,Calogero:1963,Calogero:1964,Babikov:1967,Tikochinsky:1970}.
Exhaustive treatises on the subject are found in books by Razavy \cite{Razavy:2003},
Babikov \cite{Babikov:1968}, and Calogero \cite{Calogero:1967}.
Solving differential equations for the phases of the stationary-state
wave functions, as functions of coordinate, is central to the VPM
approach. These phases at asymptotic distances (from the scattering
potential) make up the $S$-matrix of the problem. Equations for such
quantities can be found by considering the phase shifts corresponding
to the scattering potential being truncated at some coordinate locations.
Alternatively, Green's function approach can be used. Techniques of
this sort are widely used in reaction physics with atoms, molecules,
and nuclei, and in relativistic scattering. Recently, there have been
interests centered around multi-channel tunneling and scattering problems
\cite{Goodvin:2005,Balantekin:1998,Kayanuma:1994,Razavy:2003,Hnybida:2008,Shegelski:2008,Shegelski:2008i}.
Our problem is unlike those ordinarily encountered because its solution
depends on proper treatment of the multi-channel virtual dynamics.
Thus, here we enhance the VPM by applying it to virtual channels,
which is mentioned in Ref. \cite{Babikov:1968} as a possibility.
Some later steps in this direction have been taken in Ref. \cite{Talukdar:1981}
with off-shell amplitudes in the context of a three-body problem.

\subsection{Formulation of the VPM}

Let us first introduce the VPM briefly. We would like to emphasize
that though we limit our discussion to one-dimensional scattering
for simplicity, the approach is a general one. This method is also
known as the Variable Reflection Amplitude Method \cite{Kiers:1996,Razavy:2003,Tikochinsky:1977}.
By using factorization of the form \[
\Phi(X,x)=\sum_{n}\Psi_{n}(X)\psi_{n}(x)\]
for the wave function, Schrödinger's equation $H\Phi(X,x)=E_{T}\Phi(X,x)$
with the Hamiltonian from \eqref{eq:Hamiltonian}-\eqref{eq:intrinsicH}
can be transformed into a coupled-channel equation for the center-of-mass
wave-functions $\Psi_{n}(X)$ for channels $n$ (subject to appropriate
boundary conditions) \begin{equation}
\left[\frac{\partial^{2}}{\partial X^{2}}+K_{n}^{2}\right]\Psi_{n}(X)-\sum_{n'}V_{nn'}(X)\Psi_{n'}(X)=0,\label{eq:cc_SE}\end{equation}
where the folded potentials are \begin{equation}
V_{nn'}(X)=\frac{2M}{\hbar^{2}}\int_{-\infty}^{\infty}\psi_{n}^{*}(x)\, V(X,x)\,\psi_{n}(x)dx\,,\label{eq:FoldedPot}\end{equation}
and $K_{n}$ is defined in Eq. \eqref{eq:define_K}. 

The reflection and transmission amplitudes are specified in reference
to the potential-free solutions of Schrödinger's equation. These solutions
are defined in terms of diagonal matrices as 

\begin{equation}
\Xi_{nn'}^{\pm}(X)=\frac{e^{\pm iK_{n}X}}{\sqrt{-2iK_{n}}}\,\delta_{nn'},\label{eq:VPM-OPM}\end{equation}
where the $\pm$ sign corresponds to a wave moving in the right/left
direction. These solutions are normalized to unit current with the
Wronskian set to unity,\[
\Xi^{+}(X)\frac{d\Xi^{-}(X)}{dX}-\Xi^{-}(X)\frac{d\Xi^{+}(X)}{dX}=1.\]
The functions defined in \eqref{eq:VPM-OPM} can be used for both
open and closed channels provided that, as mentioned earlier, the
principal branch of the square-root is selected for an imaginary $K_{n}$. 

While there are variations of the VPM technique \cite{Razavy:2003},
we demonstrate here the approach that explicitly emphasizes the decoupling
of the reflection and transmission coefficients and the different
roles thereof \cite{Babikov:1968}.

It is convenient to apply the VPM by considering an auxiliary set
of free-space wave functions

\begin{equation}
\Psi(X,X')=\left[\Xi^{+}(X)+\Xi^{-}(X)\, R(X')\right]\overline{T}(X')\label{eq:VAM_WF}\end{equation}
with coefficients $R(X')$ and $\overline{T}(X')$ defined from the
solution $\Psi(X)$ of Schrödinger's Eq. \eqref{eq:cc_SE}, using
the Cauchy boundary condition at some point $X',$ so that at $X=X'$
\begin{equation}
\Psi(X,X')=\Psi(X)\quad\mathrm{and}\quad\frac{d}{dX}\Psi(X,X')=\frac{d}{dX}\Psi(X).\label{eq:VPM_AX}\end{equation}

\begin{figure}[!ht]
\begin{centering}
\includegraphics[angle=-90,width=8cm]{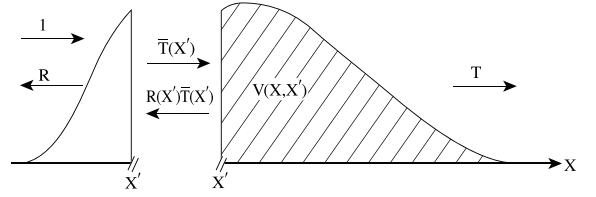}
\par\end{centering}

\caption{An incoming wave of amplitude $1$ is traveling from the left toward
the potential barrier. It is reflected with amplitude $R$ and transmitted
with amplitude $T$. For any arbitrary point $X'$ the barrier is
thought of as a combination of two parts: the unshaded part to the
left of $X'$, and the shaded part $V(X,X')$ to the right. At $X'$,
in accordance with Eqs. \eqref{eq:VAM_WF} and \eqref{eq:VPM_AX},
the incoming and outgoing components are identified as having amplitudes
$\overline{T}(X')$ and $R(X')\overline{T}(X')$ respectively. Here
$\overline{T}(X')$ represents the overall amplitude of the wave function
at $X'$ which has been modified, relative to the incoming beam, due
to the passage through the unshaded part of the barrier. In the context
of the shaded part only, $\overline{T}(X')$ represents an incident
beam normalization; thus $R(X')$ is interpreted as the amplitude
of reflection from the shaded part. Due to this normalization, the
transmission amplitude $T(X')$ through the shaded part is given by
the final amplitude $T$ (transmission through the full potential)
normalized relative to the incident amplitude $\overline{T}(X').$
Thus, $T(X')\,\overline{T}(X')=T.$ The incoming beam has a unit amplitude,
and $\overline{T}(-\infty)=1,\quad T(-\infty)=T.$ It is obvious that
$R(-\infty)=R$ which is the amplitude of reflection from the full
potential. On the right of the potential $R(\infty)=0,\quad\overline{T}(\infty)=T,$
and $T(\infty)=1.$ \label{Illustration_of_VPM}}

\end{figure}

It is convenient to interpret $\Psi(X,X')$ for $X\le X'$ as the
wave function corresponding to a potential truncated from the left,
$V_{nn'}(X,X')=V_{nn'}(X)\,\theta(X-X')$, where $\theta(X)$ is the
Heaviside step function. So, $V_{nn'}(X,X')=0$ at $X\le X'$, and
the wave function is given by (\ref{eq:VAM_WF}) with the boundary
condition (\ref{eq:VPM_AX}). This interpretation is illustrated in
Fig. \ref{Illustration_of_VPM}. Now, $R(X')$ in (\ref{eq:VAM_WF})
is a matrix in the channel space. With the help of Fig. \ref{Illustration_of_VPM},
it can be identified as the reflection amplitude for a wave scattering
from the potential truncated from left (the shaded part in the figure).
The vector $\overline{T}(X')$ in the channel space is the amplitude
of the wave function $\Psi(X,X')$ at $X=X'$. It can be normalized
in different ways. In the literature the wave function $\Psi$ and
its amplitude $\overline{T}$ are commonly viewed as collections of
independent column vectors corresponding to different independent
initial conditions (initial channels). It is clear from Fig. \ref{Illustration_of_VPM}
that transmission through the potential $V(X,X')$ is inversely proportional
to the amplitude $\overline{T}(X')$. With relatively straight-forward
derivations that involve substitution of the wave function in Schrödinger's
Eq. (\ref{eq:cc_SE}) by (\ref{eq:VAM_WF}) (see also Ref. \cite{Babikov:1968}),
one can show that the matrix $R(X)$ is subject to the differential
equation
\begin{equation}
\frac{dR(X)}{dX}=\left[\left(\Xi^{+}+R(X)\,\Xi^{-}\right)\right]V\left[\Xi^{+}+\Xi^{-}\, R(X)\right].\label{eq:VPM_R}\end{equation}
The $X$-dependency of the $\Xi$'s and the folded potential
$V$ is suppressed in this differential equation and all others to
follow unless there is an ambiguity.

The equation for the vector $\overline{T}(X)$ is linear, \begin{equation}
\frac{d\overline{T}(X)}{dX}=-\Xi^{-}\, V\,\left[\Xi^{+}+\Xi^{-}\, R(X)\right]\overline{T}(X),\label{eq:VPM_T}\end{equation}
which reflects linearity of quantum mechanics. To be more
specific, the linearity shows that the column-vectors corresponding
to different initial channels are independent of each other, and the
amplitudes allow for arbitrary normalizations.

The reaction physics of interest, in agreement with the discussion
in Sec. \ref{sub:Smatrix}, is given by $R$ alone; and, the physical
properties are independent of normalization, resulting in the decoupled
equation \eqref{eq:VPM_R} for $R(X)$. 

As follows from the boundary condition on the wave function $\Psi(X)$
or from the interpretation of $R(X')$ as a reflection amplitude,
$R(X')$ is subject to the boundary condition \begin{equation}
R_{nn'}(\infty)=0,\quad\mathrm{\mathrm{that\,}leads\, to}\quad R_{nn'}(-\infty)=R_{nn'}.\label{eq:VPM_BR}\end{equation}

The transmission amplitude is determined by $\overline{T}(X')$. One
can treat $\overline{T}(X')$ as a matrix of the column-vectors described
previously. Assuming the incident beam to be in channel $n'$ and
normalizing it to unity (which is the most common and natural way),
one would have \begin{equation}
\overline{T}{}_{nn'}(-\infty)=\delta_{nn'},\quad\mathrm{and}\,\mathrm{thus,\quad}\overline{T}_{nn'}(\infty)=T_{nn'}\,.\label{eq:VPM_BT}\end{equation}
The elements of the reflection matrix $R(X)$ are sufficient
to determine all observable probabilities. One, however, may still
want to obtain the transmission amplitudes, which can be done by integrating
Eq. \eqref{eq:VPM_T} separately using previously determined values
of $R(X)$. Due to the different boundary conditions for $R$ and
$T$ (see Eqs. \eqref{eq:VPM_BR} and \eqref{eq:VPM_BT}), this approach
is computationally inconvenient. However, given that $\overline{T}{}_{nn'}(\infty)=T_{nn'}$,
it can be interpreted as a final state normalization to a yet-unknown
value $T_{nn'}$, and this inconvenience can be avoided. If one defines
a matrix $T(X)$ so that $T(X)\overline{T}(X)=T$. Then $T(X)$ coincides
exactly with the matrix of transmission amplitudes through the truncated
potential $V(X,X')$ (see Fig. \ref{Illustration_of_VPM}). Since
$\frac{d}{dX}(T(X)\overline{T}(X))=0$, Eq. \eqref{eq:VPM_T} in terms
of $T(X)$ is \begin{equation}
\frac{dT(X)}{dX}=T(X)\,\Xi^{-}\, V\,\left[\Xi^{+}+\Xi^{-}\, R(X)\right].\label{eq:VPM_T-1}\end{equation}
It is to be used with the boundary condition \begin{equation}
T_{nn'}(\infty)=\delta_{nn'},\quad\mathrm{and\: then}\quad T_{nn'}(-\infty)=T_{nn'}\,.\label{eq:VPM_BT-1}\end{equation}
This approach is equivalent to the one discussed in Ref. \cite{Razavy:2003}.

In summary, the most celebrated advantages of the VPM are its physical
transparency, generality, and the simplicity in its application. Phase
equations can indeed be constructed for most quantum mechanical problems.
In particular, with appropriate substitutions for the $\Xi^{\pm}$
functions, the approach can be immediately used in three dimensional
problems with radial variables. The VPM is technically simple; the
entire multi-channel problem is reduced to a relatively straight-forward
integration of Riccati equation \eqref{eq:VPM_R} from right to left
with a zero starting-value (for $R(X)$) as a boundary condition.

\subsection{Virtual channels in the VPM}

In this work we find yet another value of the VPM in its effectiveness
in treating virtual channels and in general complex-momentum (i.e.,
off-shell) applications. Some suggestions, in this direction, have
been made in Refs. \cite{Babikov:1968,Talukdar:1981}. However, there
are a few important things to note:

\emph{First}, the formalism remains valid in the complex momentum
plane assuming that for virtual channels the principal branch of the
square root is selected in Eq. \eqref{eq:define_K}. 

\emph{Secondly}, separation of the amplitude (given by $\overline{T}(X)$)
from the physically relevant phase difference between incoming and
outgoing components (given by $R(X)$) is important. Due to this separation,
the principal equation \eqref{eq:VPM_R} is solved with a zero-value
boundary condition \eqref{eq:VPM_BR}, and without any concern about
exponentially falling or rising components of the wave function outside
the potential. Normalization is provided by a set of decoupled equations
for each selected initial condition. Therefore, while solving scattering
problems, one could consider only those columns $\overline{T}(X)$
that correspond to open channels of interest.

Closed channels can be studied, if desired, with an initial wave function
exponentially rising toward the potential. Bound states can also be
explored in this way \cite{Calogero:1967}, but we do not study these
questions.

Normalizing in a way to have the closed channels set to zero still
deserves some attention. As further demonstrated in Sec. \ref{sub:Spatial-dynamics},
$T(X)$ and $\overline{T}(X)$ for closed channels are set to zero
differently. For $\overline{T}(X)$ one assumes the initial beam normalization
of zero for any closed channel, i.e.,
\begin{equation}
\overline{T}_{nn'}(-\infty)=0\qquad\mathrm{if\:}n'\:\mathrm{is\: closed}.\label{eq:VPM_BC_Tb_virtual}\end{equation}
Thus after scattering one has waves in closed channels with
amplitudes exponentially decaying to zero away from the potential. 

Contrary to that, eq. \eqref{eq:VPM_BT-1} is best thought of as a
final state normalization, thus
\[
T_{nn'}(\infty)=0\qquad\mathrm{\textrm{if\:\ n\:\ is\:\ closed}}.\]

\emph{Finally}, the original Eqs. \eqref{eq:VPM_R} and \eqref{eq:VPM_T}
used for open channels, are valid also for closed channels, but have
issues with numerical stability for virtual excitations. The exponential
divergence of the functions $\Xi_{nn}^{\pm}(X)$ with $n$ (or, $|K_{n}|$),
especially at large $|X|$, makes it difficult to handle long-ranged
potentials. We define \[
U(X)=-\Xi^{+}\Xi^{-}-\Xi^{-}\, R(X)\,\Xi^{-},\]
which agrees with Eq. \eqref{eq:wall-U}, so that Eq. \eqref{eq:VPM_R},
written in terms of the variables $U(X)$ instead of $R(X)$, reads
\begin{equation}
\frac{dU_{nn'}}{dX}=\delta_{nn'}-i(K_{n}+K_{n'})U_{nn'}-\sum_{ll'}U_{nl}V_{ll'}U_{l'n'},\label{eq:VAM-U}\end{equation}
where $U$ and $V$ depend on $X$. It is noteworthy that
in this form the equations no longer contain any exponential factors.
A similar substitution can be done for $T(X)$ or $\overline{T}(X)$. 

The amplitudes $U$ for open channels oscillate outside the potential
where $V=0$, which is not the most desirable boundary condition one
would want to deal with. However, this is a minor inconvenience compared
to the benefit of the exponential drop of $U(X)$ for virtual channels
with distance from the potential barrier. This is particularly important
because in the problems that we discuss, the reaction processes contain
only a few open channels but are determined by numerous closed channels.

\section{Applications of the VPM \label{sec:Applications}}

\subsection{A $\delta$-barrier\label{sub:A delta-barrier}}

We proceed by considering a $\delta$-barrier as the scattering potential.
Here, a bound system of two particles is incident on a potential \begin{equation}
V(x_{1},\, x_{2})=\frac{\hbar^{2}}{AM}\delta(x_{2}),\label{eq:delta}\end{equation}
where, again, only the second particle interacts with the potential;
$A$ is a length parameter, characterizing the strength of the barrier. 

Interactions of composite objects, such as diatomic molecules, with
a $\delta$-barrier have been discussed before \cite{Razavy:2003,Shegelski:2008,Hnybida:2008},
but usually without any involvement of the virtual channels and in
situations where the potential-barrier acts on both the particles.

Any potential can be considered with the VPM approach in principle;
however, the short-ranged $\delta$-potential provides a good way
of exploring the generic features of scattering without putting efforts
into computing folded-potentials. The folded potential \eqref{eq:FoldedPot}
for a $\delta$-barrier \eqref{eq:delta} takes an analytic factorized
form: \begin{equation}
V_{nn'}(X)=\frac{2}{\mu_{1}A}\;\psi_{n}^{*}\left(\frac{X}{\mu_{1}}\right)\;\psi_{n'}\left(\frac{X}{\mu_{1}}\right),\label{eq:DeltaFoldedPot}\end{equation}
where $\mu_{1}$ is the mass-ratio defined in \eqref{eq:mass-ratios}.

In the limit of $A\rightarrow0$ the $\delta$-barrier turns into
an impenetrable wall, thus allowing us to complete the study in Sec.
\ref{truncation}, which is done in Sec. \ref{sub:Infinite-wall}.

Introduction of the short-ranged $\delta$-barrier adds just one additional
length scale $A$ to the parameters used to describe the problem.
There are thus three length scales in the problem, namely, the intrinsic
scale $\lambda$, the incident beam wavelength $\sim1/K$, and the
potential scattering length $A.$ The corresponding energy scales
are the intrinsic energy scale $\epsilon$, the incident beam kinetic
energy $E$, and the energy scale associated with the $\delta$-potential,
defined by

\begin{equation}
E_{\delta}=\frac{\hbar^{2}}{2MA^{2}}.\label{eq:Edelta}\end{equation}

The mass-ratio $\mu$ defined in Eq. \eqref{eq:mass-ratios} connects
the length scales $\lambda$ and $1/K$ at similar energies; the precise
relation is $\lambda K=\sqrt{E/\mu}$.

The non-composite limit of the process is reached either if all the
mass is concentrated in the interacting particle leaving $\mu_{1}=0$
and therefore $\lambda\rightarrow\infty$, or if the intrinsic states
have infinitely high energy $\epsilon\rightarrow\infty$ and thus
$\lambda=0$. This yields a textbook problem of scattering off a $\delta$-barrier,
where the transmission and reflection amplitudes are, respectively,
\begin{equation}
T=\frac{iKA}{iKA-1},\quad\mathrm{and\quad}R=\frac{1}{iKA-1}.\label{eq:deltaamplitude}\end{equation}
For a non-composite projectile the $\delta$-potential allows for
scattering only in the symmetric channel, since \[
S^{+}=\frac{1+iKA}{1-iKA},\quad\mathrm{and\quad}S^{-}=1.\]
Transmission and reflection probabilities from a $\delta$-barrier
are determined solely by the energy ratio $(KA)^{2}=E/E_{\delta}$.
Thus, the sign of the coupling $A$, i.e., whether it is a well or
a barrier, does not matter,\begin{equation}
|T^{2}|=\frac{E/E_{\delta}}{E/E_{\delta}+1}\,,\quad|R^{2}|=1-|T^{2}|.\label{eq:deltaRT}\end{equation}

\subsubsection{Spatial dynamics of the reflection and transmission amplitudes \label{sub:Spatial-dynamics}}

\begin{figure}[!ht]
\begin{centering}
\includegraphics[width=7cm]{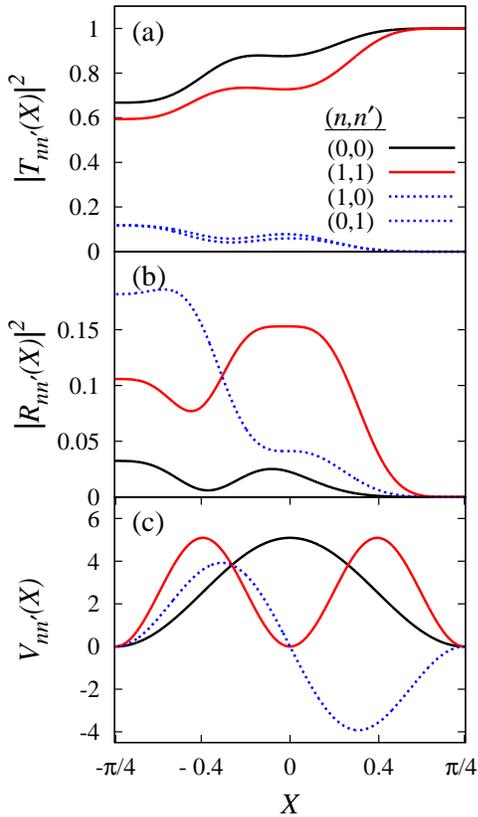}
\par\end{centering}

\caption{(Color online) \textbf{Well and $\delta$-barrier.} Barrier strength
is $E_{\delta}=1$. With incident beam kinetic energy $E=4$, the
only open channels are those labeled by $0$ or $1$. The transmission
and reflection probabilities, $|T_{nn'}(X)|^{2}$ and $|R{}_{nn'}(X)|^{2}$,
and the folded potentials $V_{nn'}(X)$ are shown in panels (a), (b)
and (c) for the open channels. $X$ is expressed in units of $\lambda$.\label{ddynamic}}

\end{figure}

\begin{figure}[!ht]
\begin{centering}
\includegraphics[width=7cm]{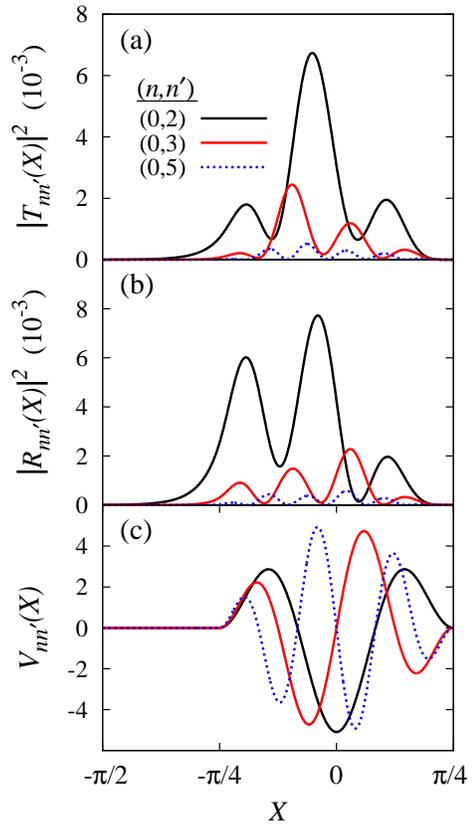} 
\par\end{centering}

\caption{(Color online) \textbf{Well and $\delta$-barrier.} This graph refers
to the same projectile as described in Fig. \ref{ddynamic}, with
$E=4$. The reflection and transmission cross sections, $|R_{nn'}(X)|^{2}$
and $|T_{nn'}(X)|^{2}$, the folded potentials $V{}_{nn'}(X)$ are
shown for the out-going channel, $n=0$, for a projectile in three
virtual incoming channels $n'=2,\:3,\:5$. $X$ is expressed in units
of $\lambda$. \label{vdynamic}}

\end{figure}

The spatial dynamics of the reflection and transmission amplitudes
is shown in Figs. \ref{ddynamic} and \ref{vdynamic}. As an example
we take the \textquotedbl{}well\textquotedbl{} confinement (see Sec.
\ref{sub:Infinite-square-well}) with equal particle masses, $\mu_{1}=\mu_{2}=1/2,$
and a $\delta$-barrier with strength $E_{\delta}=1$ in units of
intrinsic excitations \eqref{eq:energyunit}. The kinetic energy of
the beam is $E=4$ in the same units, which means that there are two
open channels.

In each of these figures, the lower panel (c) shows folded potentials,
as follow from Eqs. \eqref{eq:DeltaFoldedPot} and \eqref{eq:eigenSQWellPot}.
Thanks to the simplicity of $\delta$-barrier; the folded potentials
have obvious forms showing the structures of the wave functions for
the intrinsic square well confinement. Naturally, $V_{nn'}(X)=0$
outside the well, or, in other words, if $|{x}|\ge\pi/2$ (that is,
$|X|\ge\pi/4$) where $x$ (as well as $X$) is expressed in units
of $\lambda$.

As explained through Fig. \ref{Illustration_of_VPM}, the dynamic
transmission and reflection amplitudes, $T(X)$ and $R(X)$, in the
VPM correspond to the potential truncated from the left of $X$. Therefore,
both transmission and reflection probabilities $|T_{nn'}(X)|^{2}$
and $|R_{nn'}(X)|^{2}$ shown in panels (a) and (b), respectively,
are evolved from right to left following Eqs. (\ref{eq:VPM_BR}) and
(\ref{eq:VPM_BT-1}). For both Figs. \ref{ddynamic} and \ref{vdynamic}
we utilize the final state normalization (\ref{eq:VPM_BT-1}) of $T(X)$
instead of $\overline{T}(X)$, due to the convenience in application
and interpretation of $T$ as the transmission amplitude.

Figure \ref{ddynamic} shows the dynamics in the open channels. Since
the intrinsic potential is of a finite width, the final values of
the reflection and transmission coefficients are reached at $X=-\pi/4$
which means inclusion of the full potential. And therefore, the values
of $|R_{nn'}(X)|^{2}$ and $|T_{nn'}(X)|^{2}$ at $X=\pm\pi/4$ are
the asymptotic values thereof, commensurate with \eqref{eq:VPM_BR}
and \eqref{eq:VPM_BT-1}. The probability is conserved at all values
of $X$: \[
\sum_{n\in\mathrm{open}}|T_{nn'}(X)|^{2}+|R_{nn'}(X)|^{2}=1,\]
and $R_{nn'}(X)=R_{n'n}(X)$ due to time reversal symmetry. However,
$T_{nn'}(X)\ne T_{n'n}(X)$ due to the asymmetry in the truncated
potential. Symmetry is recovered in the final $T$ since the full
symmetric potential is covered at $X=-\pi/4$, where $|T_{nn'}|^{2}=|T_{n'n}|^{2}$,
as seen in Fig. \ref{ddynamic}(a).

A different picture emerges with the virtual channels. The linearity
and independence of initial conditions, discussed earlier, is important
since it makes the normalization of virtual channels irrelevant for
the S-matrix and other asymptotic reaction observables. Having said
that, one can assume that $\overline{T}_{nn'}(-\infty)=0$ if the
initial channel $n'$ is closed (Eq. \eqref{eq:VPM_BC_Tb_virtual}),
and thus $\overline{T}_{nn'}(X)=0$ for any $X$ due to linearity
(\ref{eq:VPM_T}). However, an incident beam in some open channel
$n'$ generates virtual excitations $n$ that exist outside the potential.
Therefore, $\overline{T}{}_{nn'}(X)$ is an exponentially decaying
non-zero function beyond the range of the potential when initial channel
$n'$ is open but the final channel $n$ is closed. 

A totally different situation arises with the final state normalization
\eqref{eq:VPM_BT-1}, i.e., with $T_{nn'}(\infty)=0$ if $n$ corresponds
to a closed channel. Thus assuming that all virtual channels are normalized
to zero to the right of the barrier generates disturbances of virtual
channels in front of the potential barrier, so that $T_{nn'}(X)$
is not zero there when $n'$ is closed. This is seen in Fig. \ref{vdynamic}
where such quantities $T_{nn'}(X)$ (or, their norms) decay exponentially
to the left of the barrier, but are not zero at $X=-\pi/4$. 

The interpretation of $R_{nn'}(X)$ is quite different. It sets relations
between the different components (phase shifts) of the wave function,
and is non-zero for all real and virtual initial and final channels.
Nevertheless, the behavior is similar (see Fig. \ref{vdynamic}).

\subsubsection{Results\label{sub:Results}}

Let us now discuss some final results for the scattering and tunneling
of the deuteron-like system. Here we continue to consider the infinite
square well ({}``well'') and the harmonic oscillator ({}``HO'')
models (see Sec. \ref{sub:Infinite-square-well} and \ref{sub:Harmonic-oscillator-confinement}),
that do not allow for breakup. A model with a continuum of intrinsic
states, which allows for breakup, is described in Sec. \ref{sec:continuum}.

\begin{figure}
\includegraphics[width=7cm]{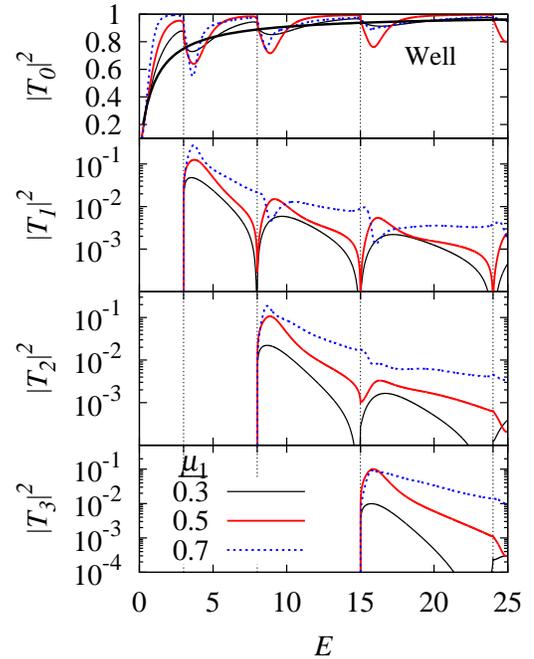}\caption{(Color online) \textbf{Well and $\delta$-barrier.} Barrier strength
$E_{\delta}=1$. Probabilities of transmission to different final
channels $n=0,\,1,\,2\:\textrm{and}\:3$ are shown as functions of
incident beam kinetic energy $E$. The incident beam is in the ground
state channel. The different curves on each panel correspond to different
values of the mass-ratio $\mu_{1}$, as labeled. The non-composite
limit is shown in the top panel with the thick solid line. \label{fig:VRA-SW_T}}

\end{figure}

\begin{figure}
\includegraphics[width=7cm]{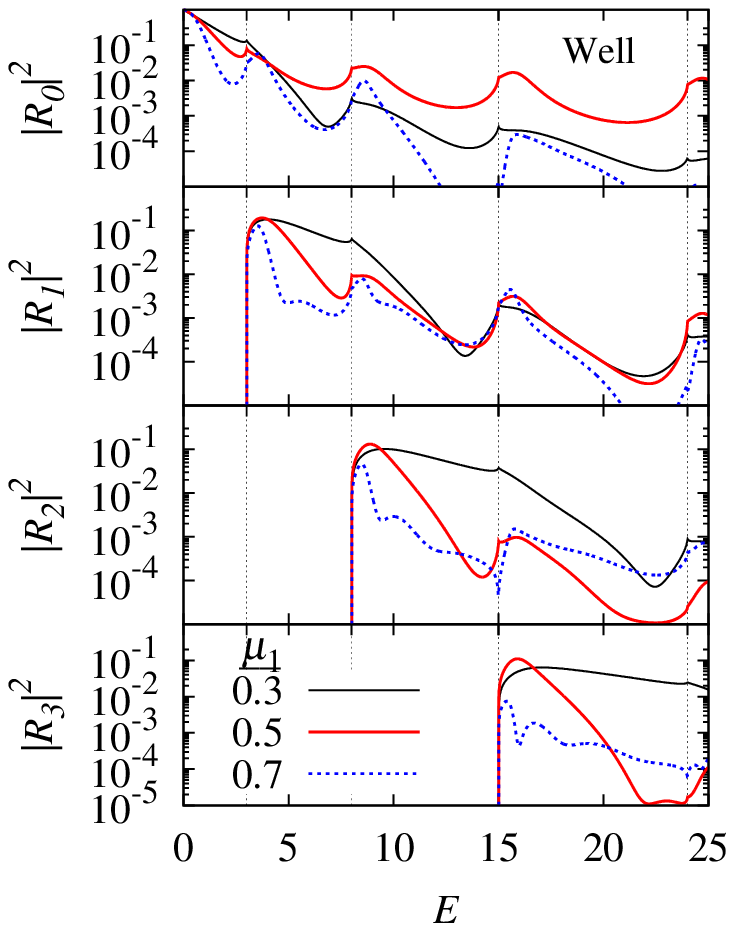}\caption{(Color online) \textbf{Well and $\delta$-barrier.} Same as Fig. \ref{fig:VRA-SW_T},
showing reflection probabilities.\label{fig:VRA-SW_R}}

\end{figure}

\begin{figure}
\includegraphics[width=7cm]{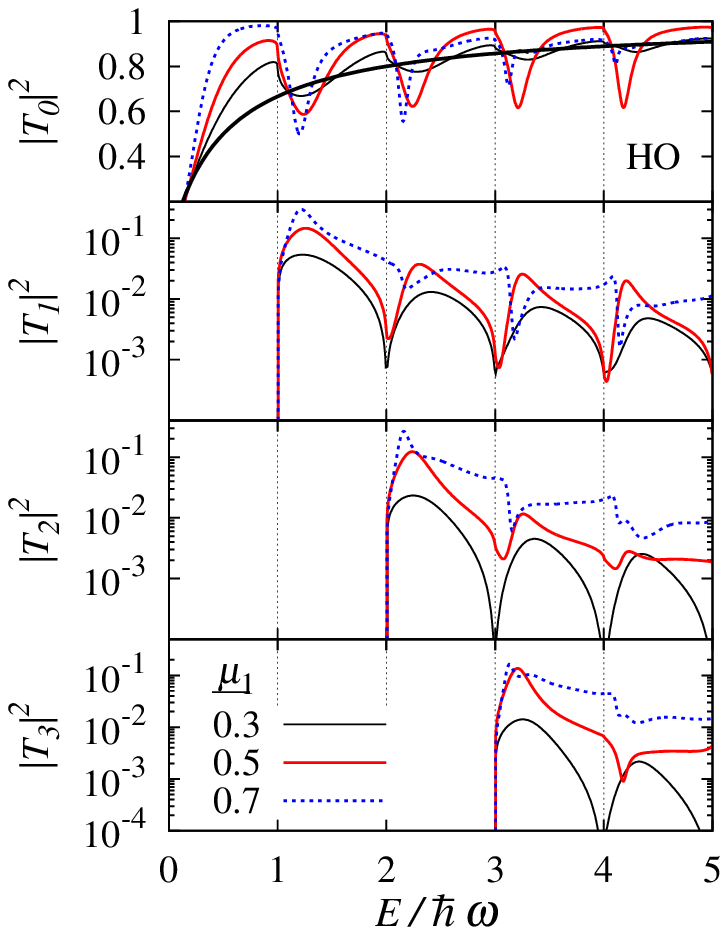}\caption{(Color online) \textbf{HO and $\delta$-barrier.} Barrier strength
$E_{\delta}=1=0.5\:\hbar\omega$. Probabilities of transmission to
different final channels $n=0,\,1,\,2\:\textrm{and}\:3$ are shown
as functions of incident beam kinetic energy $E$. The incident beam
is in the ground state channel. The different curves on each panel
correspond to different values of the mass-ratio $\mu_{1}$, as labeled.
The non-composite limit is shown in the top panel with the thick solid
line. \label{fig:VRA-HO_T}}

\end{figure}

\begin{figure}
\includegraphics[width=7cm]{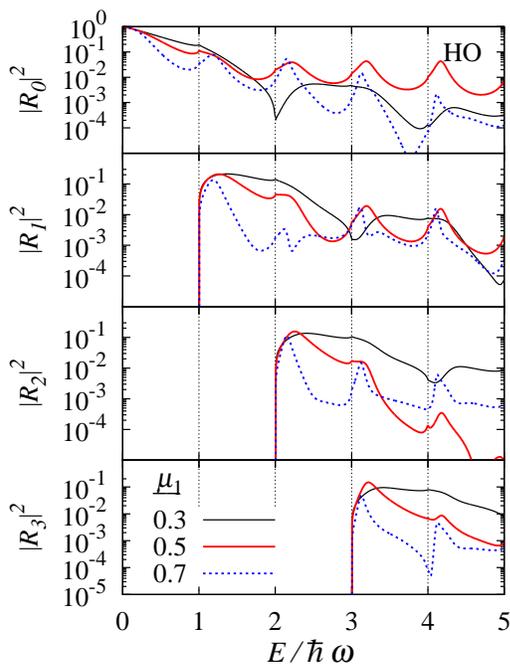}\caption{(Color online) \textbf{HO and $\delta$-barrier.} Same as Fig. \ref{fig:VRA-HO_T},
showing reflection probabilities.\label{fig:VRA-HO_R}}

\end{figure}

Figures \ref{fig:VRA-SW_T} and \ref{fig:VRA-SW_R} for the {}``well,''
and Figs. \ref{fig:VRA-HO_T} and \ref{fig:VRA-HO_R} for the {}``HO,''
depict the probabilities $|T_{n}|^{2}$ of transmission and $|R{}_{n}|^{2}$of
reflection, respectively, in the first few lowest channels as functions
of incident beam kinetic energy $E$. Each figure contains vertical
grid lines indicating the locations of channel thresholds. In the
case of the \textquotedbl{}well,\textquotedbl{} new channels open
up at kinetic energies $E=n^{2}-1$ where $n$ is a positive integer;
for the \textquotedbl{}HO,\textquotedbl{} these occur at integral
multiples of $\hbar\omega$. In each case the incident beam is in
the ground state channel, thus the corresponding subscript is suppressed,
and $n$ refers to the final channel. The strength of the $\delta$-barrier
is set, via Eq. \eqref{eq:Edelta}, to $E_{\delta}=1$. The redistribution
of probabilities at the threshold energies, required by unitarity,
leads to cusps in the cross sections \cite{Landau:1981,Baz:1971}.
These discontinuities are common in all the figures for both the models. 

All figures contain three curves with $\mu_{1}=0.3,\:0.5\:\mathrm{and}\:0.7$,
and thus illustrates the mass-ratio dependence. In addition to that
are shown the ground state to ground state transmission probabilities
in the top panels of Figs. \ref{fig:VRA-SW_T} and \ref{fig:VRA-HO_T}
in the non-composite limit $\mu_{1}\to0$ where an analytic answer
follows from Eq. \eqref{eq:deltaamplitude}. Indeed, when the non-interacting
particle-1 is very light compared to the interacting particle-$2$,
i.e., when $\mu_{1}\rightarrow0$, then particle-$2$ carries almost
all the momentum, and the presence of particle-$1$ hardly matters.
In this limit the behavior of the projectile approaches that of a
single non-composite particle. 

The development of the resonant behavior, as $\mu_{1}$ increases,
is easy to follow in these plots. For larger values of $\mu_{1}$
the curves exhibit prominent peaks and dips that are not associated
with cusps at thresholds. Classically this can be viewed as a process
in which the light interacting particle is stopped by the potential,
while the larger mass $\mu_{1}$ keeps moving forward without any
impediment, until most of its kinetic energy is transferred into potential
energy of the intrinsic interaction, and then it either turns back
or pulls the smaller interacting mass through the barrier. Hence,
the larger the non-interacting particle's mass, the more complex and
chaotic the process.

\begin{figure}
\includegraphics[width=7cm]{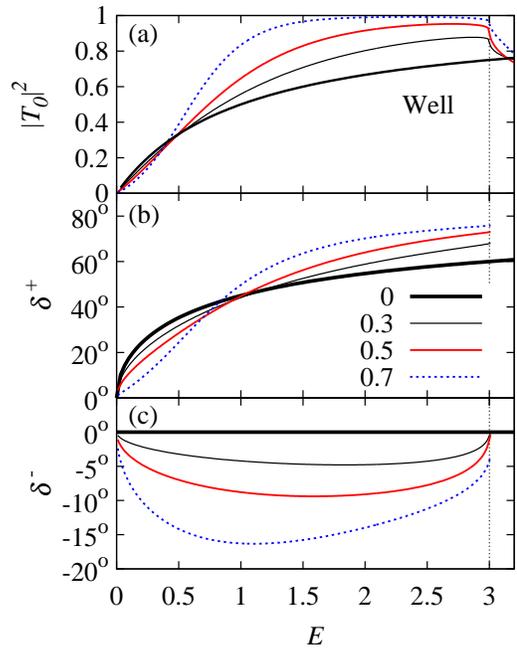}\caption{(Color online) \textbf{Well and $\delta$-barrier.} The system is
same as that in Fig. \ref{fig:VRA-SW_T}, showing transmission probability
$|T_{0}|^{2}$ for energies below the first threshold in panel panel
(a), and phase shifts $\delta^{\pm}$ for the same energy region in
panels (b) and (c). \label{fig:VRA-SW_phase}}

\end{figure}

\begin{figure}
\includegraphics[width=7cm]{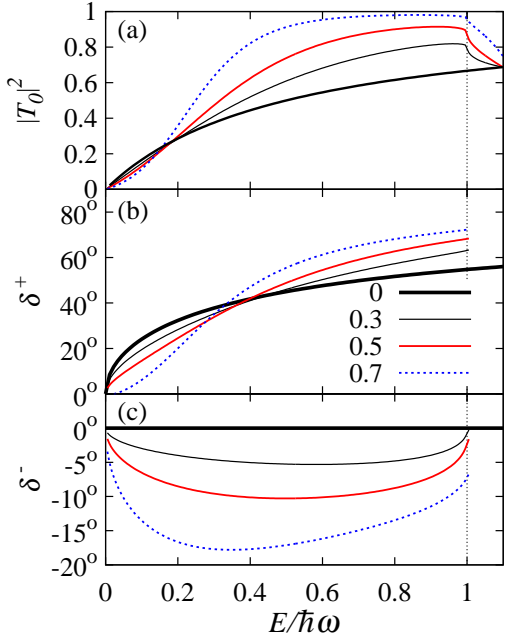}\caption{(Color online) \textbf{HO and $\delta$-barrier.} The system is same
as that in Fig. \ref{fig:VRA-HO_T}, showing transmission probability
$|T_{0}|^{2}$ for energies below the first threshold in panel panel
(a), and phase shifts $\delta^{\pm}$ for the same energy region in
panels (b) and (c). \label{fig:VRA-HO_phase}}

\end{figure}

It is intuitive to suggest that the highly virtual channels have little
or no effect on observables at low-energies. It is proved otherwise
in our studies. In Figs. \ref{fig:VRA-SW_phase} and \ref{fig:VRA-HO_phase}
we focus on the low energy region, below the first threshold, for
the \textquotedbl{}well\textquotedbl{} and the \textquotedbl{}HO\textquotedbl{}
models, respectively. Here we use the same parameters as in Figs.
\ref{fig:VRA-SW_T}-\ref{fig:VRA-HO_R}, and present our results for
the non-composite limit $\mu_{1}=0$ as well as for projectiles with
$\mu_{1}=0.3,$ 0.5, and 0.7. Along with the transmission probability,
we show results for the two phase shifts (\ref{eq:define_phases})
that are defined up to the first threshold (shown by the vertical
grid line). We conclude that the compositeness, and the composition
of the projectile given by the mass-ratio of the components, are consequential
factors that determine the observables. In these models, as well as
in the ones with breakup (discussed in Sec. \ref{sec:continuum}),
we find a systematic enhancement of tunneling probability, with increasing
mass of the non-interacting component, in a broad region of energy
near the first threshold. This enhancement was earlier discussed in
Ref. \cite{Ahsan:2007}. This is supported by a recent experimental
at GANIL by Lemasson and others \cite{Lemasson:2009} that shows
enhancement in tunneling of heavy He isotopes, where additional spectator-neutrons
contribute to the mass of the non-interacting component, while the
alpha core interacts with the Coulomb barrier. 


\subsubsection{An attractive $\delta$-well}

In addition to the $\delta$-barrier discussed so far, we explored
scattering that involves an attractive $\delta$-well. We stress again
that for a non-composite projectile the sign of the interaction does
not effect the observed reflection and transmission probabilities
(see Eq. (\ref{eq:deltaRT})). This is not true for a composite projectile.
This topic has been extensively explored in the literature and is
often referred to as the Barkas Effect. In Ref. \cite{Volya:2002},
one can find more references that are relevant, and a model that is
similar in spirit and discusses the Coulomb excitation of a harmonic
oscillator.

\begin{figure}
\includegraphics[width=7cm]{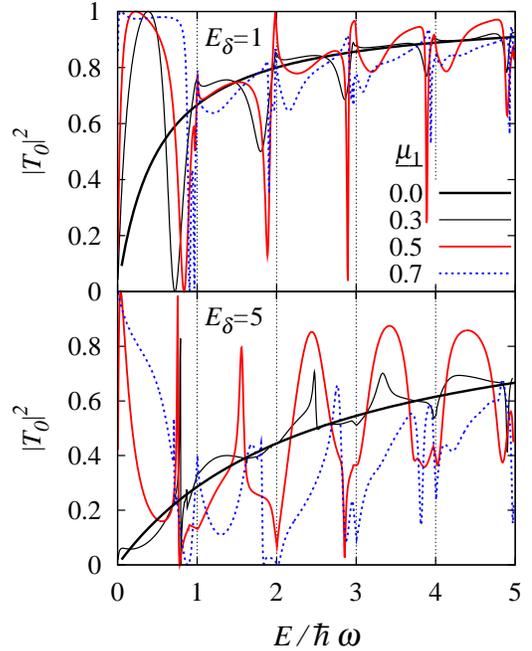}\caption{(Color online) \textbf{Well and attractive $\delta$-well.} The probability
of transmission from ground state to ground state is shown as a function
of incident kinetic energy. The upper and lower panels correspond
to the barrier-strengths $E_{\delta}=1$ and 5 (i.e., 0.5$\,\hbar\omega$
and 2.5$\,\hbar\omega$), respectively. In both cases three different
mass-ratios $\mu_{1}=0.3,$ 0.5, and 0.7 are considered, along with
the analytic limit of a non-composite projectile (labeled by $\mu_{1}=0$).
\label{fig:VRA-SW_attractive}}

\end{figure}

Our results for the transmission probability in a scattering that
involves a $\delta$-well are shown in Fig. \ref{fig:VRA-SW_attractive}.
We present results for the case of a harmonic-oscillator confinement
only; the results for the square well are similar. In all cases, even
at relatively small masses of the non-interacting component, the scattering
process is highly resonant. The interacting particle-2 and the barrier
form a bound state at an energy $-\mu_{2}E_{\delta}$, which is only
a virtual binding in the three-body problem. However, the system in
an excited state $n$ with intrinsic energy $\varepsilon_{n}$ can
be temporarily bound as a whole, thus leading to a resonance at $E_{T}=\varepsilon_{n}-\mu_{2}E_{\delta}$.
We find that this crude interpretation unravels some of the complex
resonant patterns seen in Fig. \ref{fig:VRA-SW_attractive}. The resonances
indeed periodically follow the channel thresholds, and they are close
to the thresholds for small $E_{\delta}=1$ (upper panel) and are
further away for the larger $E_{\delta}=5$ (lower panel).

\subsection{An infinite wall \label{sub:Infinite-wall}}

In this section we would like to return to the wall problem which,
as already shown in Sec. \ref{truncation}, is an extraordinarily
illustrative example. This model emerges in the limit of a very strong
$\delta$-potential (Sec. \ref{sub:A delta-barrier}), i.e., with
$A\rightarrow0$. 

\begin{figure}
\includegraphics[width=7cm]{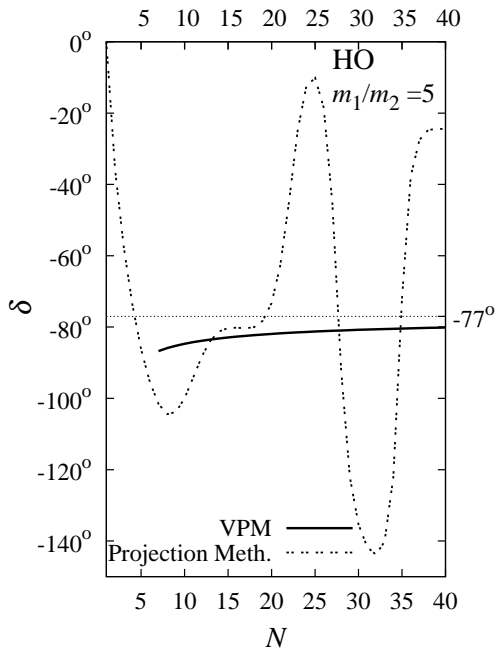}\caption{(Color online) \textbf{HO and wall.} This figure refers to the scattering
problem described in Fig. \ref{fig:osc_e05_low}, where the system
of two particles bound by \textquotedbl{}HO\textquotedbl{} confinement
collides with an infinite wall. The incident kinetic energy is $E=1=0.5\,\hbar\omega$,
and the mass-ratio is $m_{1}/m_{2}=5$. The phase $\delta$, as calculated
using the VPM, is plotted in the solid line against the number of
included channels $N$. The horizontal grid line indicates the final
value of the phase shift to which it is found to converge smoothly
with increasing $N$. The dashed curve shows results obtained through
the Projection Method, which is unstable. \label{fig:vra_conv}}

\end{figure}

We study the convergence of the VPM method separately in Sec. \ref{sec:Role-of-virtual};
nevertheless, here we present Fig. \ref{fig:vra_conv} which, in contrast
to Fig. \ref{fig:osc_e05_low}, shows that the VPM method is not prone
to the convergence issues. Even for the large mass-ratio $m_{1}/m_{2}=5$
the VPM produces a perfectly smooth curve that converges to a final
$\delta=-77{}^{o}$, which is not the case with the Projection Method. 

\begin{figure}
\includegraphics{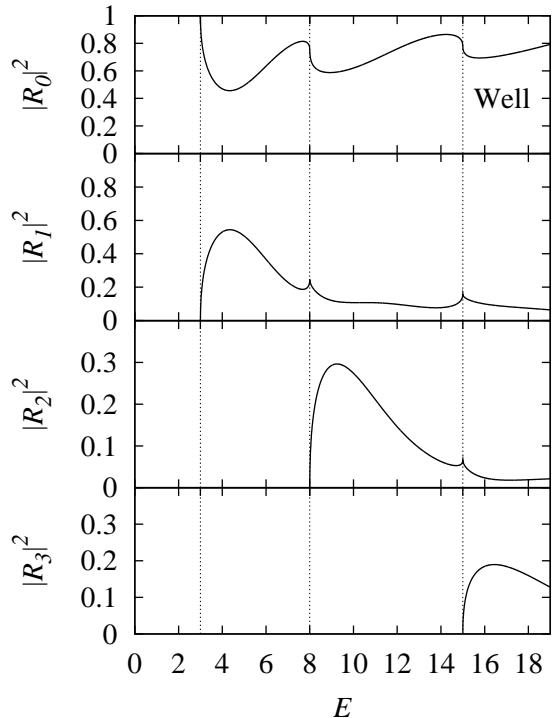}\caption{\textbf{Well and wall. }This is the same as Fig. \ref{fig:VRA-SW_R},
but for a {}``wall,'' not for a $\delta$-barrier, and only for
the case of equal masses, $m_{1}=m_{2}$ (i.e., $\mu_{1}=0.5$). \label{fig:wall-well}}

\end{figure}

\begin{figure}
\includegraphics{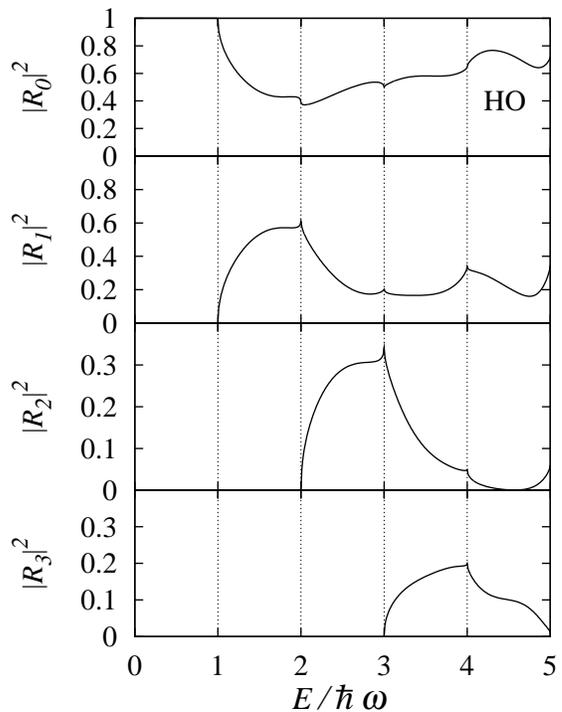}\caption{\textbf{HO and wall. }This is the same as Fig. \ref{fig:VRA-HO_R},
but for a {}``wall,'' not for a $\delta$-barrier, and only for
the case of equal masses, $m_{1}=m_{2}$ (i.e., $\mu_{1}=0.5$). \label{fig:wall-oscillator}}

\end{figure}

Figures \ref{fig:wall-well} and \ref{fig:wall-oscillator} show the
reflection probabilities of a composite projectile in the ground state
channel scattered from a wall. They are similar to the previous results
for scattering that involves a $\delta$-barrier (see Sec. \ref{sub:Results});
cusps at thresholds and some resonant behavior are among the typical
features.

\begin{figure}
\includegraphics[width=7cm]{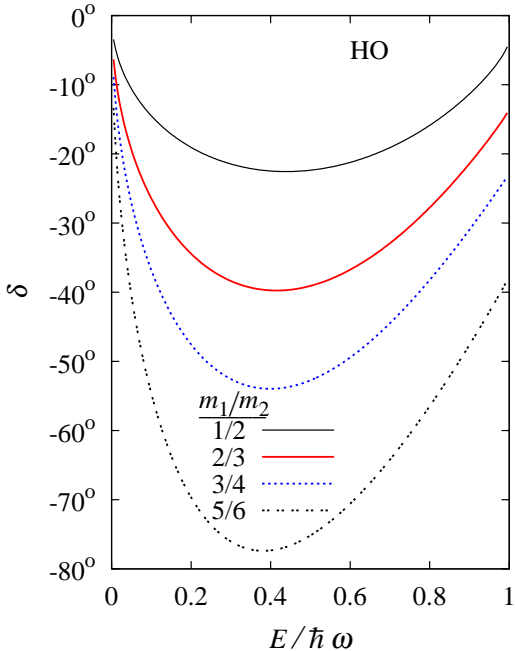}\caption{(Color online) \textbf{HO and wall. }Phase shift as a function of
incident beam kinetic energy. The curves are labeled with different
values of the mass-ratio $m_{1}/m_{2}$.\label{fig:ophase}}

\end{figure}

Scattering below the first threshold is characterized by a single
phase shift $\delta$, which is plotted in Fig. \ref{fig:ophase}
as a function of incident beam kinetic energy, in the case of the
harmonic oscillator. The different curves correspond to different
mass-ratios.

\begin{figure}
\includegraphics[width=7cm]{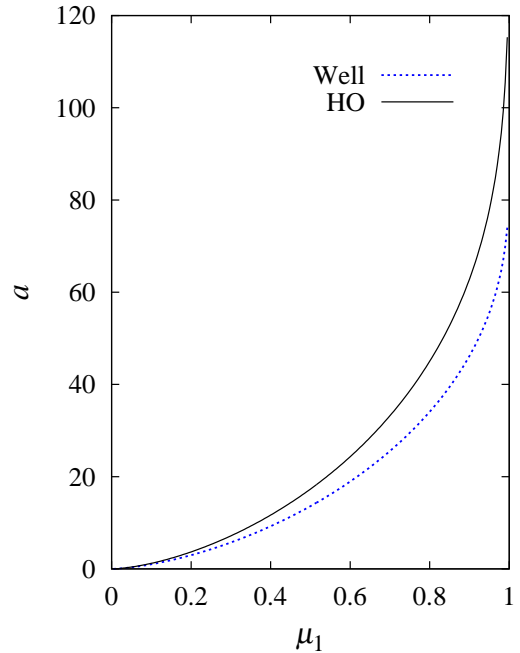}\caption{(Color online) \textbf{Well/HO and wall. }For the {}``deuteron and
Coulomb wall'' model, the scattering length $a$ (in units of intrinsic
length $\lambda$) is shown as a function of the mass-ratio $\mu_{1}=m_{1}/M$
for two different systems: those bound by the {}``well'' and by
the {}``HO'' confinements.\label{fig:olength}}

\end{figure}

The limit of very low energies is particularly interesting. The formal
effective range expansion \cite{Levy:1963,Dashen:1963,Calogero:1967,Babikov:1968},
in the context of the VPM, has been applied extensively to problems
of nucleon, molecular, and atomic scattering. As $K\rightarrow0$,
the $S$-matrix, $S=e^{2i\delta}$, is characterized by a phase $\delta=-Ka$
where $a$ is the scattering length. This length $a$ depends only
on the mass-ratio $m_{1}/m_{2}$ and represents the distance of the
turning point from the reflecting wall. A scattering length $a>0$
implies that the system is reflected at a distance $a$ prior to reaching
the wall. Figure \ref{fig:olength} shows $a$ in units of $\lambda$,
as a function of $\mu_{1}$. The limit $\mu_{1}\to0$ corresponds
to a non-composite case where the scattering length is zero. It is
interesting to note that, while the intrinsic wave function of an
infinite square well confinement has a finite width $\sim\pi\lambda$,
the scattering length can easily exceed this range. Thus, a classically
impossible situation occurs in which a finite-size system reflects
from a wall before it actually approaches it within the contact distance.
In the limit of $\mu_{1}\rightarrow$1 the scattering length ${a}$
diverges. This is a strong divergence since it is relative to a divergent
scale, $\lambda\rightarrow\infty$ for any given energy because of
the vanishing reduced mass. It is worth pointing out that the divergence
of the scattering length due to intrinsic degrees of freedom coupling
to the reaction dynamics is known as Feshbach Resonance. 

For both square well and oscillator models one can examine the analytic
results for ${a}$ by considering a few virtual channels within the
Projection Method. It becomes immediately clear that such an expansion
is convergent only in the limit of $\mu_{1}\rightarrow0$. In this
limit we obtain ${a/\lambda}\approx0.56\,\mu_{1}^{3/2}$ for the square
well bound system, and ${a/\lambda}=\mu_{1}^{3/2}$ for the oscillator-bound
system.

\begin{figure}
\includegraphics[width=6cm]{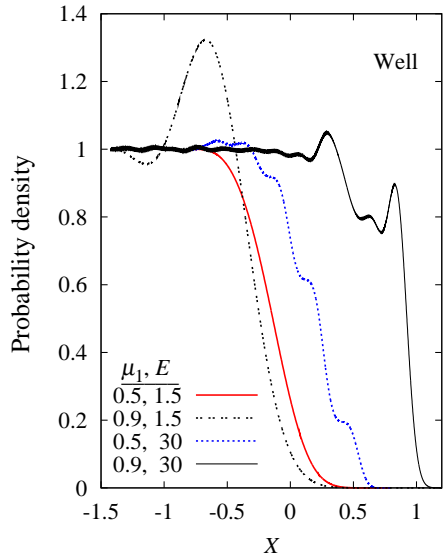}\caption{(Color online) \textbf{Well and wall. }Probability densities as functions
of location, for scattering of a square-well-bound system off an infinite
wall at $X=0$. \label{fig:wall_prob}}

\end{figure}

Figure \ref{fig:wall_prob} shows the square of the amplitude of the
wave function, $\underset{n}{\Sigma}|\overline{T}_{nn'}(X)|^{2}$,
for a projectile in the incoming channel $n'=0$. This is interpreted
as the density of probability for the center of mass of the projectile
to be at a location $X$ when it is reflected from an infinite wall.
The four curves show a few of the most representative situations;
incident beam kinetic energies as low as $E=1.5$ and as high as 30,
and two different mass-ratios $\mu_{1}=0.5$ and 0.9. All probabilities
eventually die to zero beyond the wall located at $X=0$. The first
two curves represent cases where the energy of the projectile, $E=1.5$,
is halfway between the energies of the ground state and the first
excited state. Hence only one open channel is present. For mass-ratio
$\mu_{1}=0.5$ the behavior is plain. However, when the non-interacting
particle contains 90\% of the total mass there is a peak of probability
density in front of the wall. This is consistent with the enhanced
scattering length (see Fig. \ref{fig:olength}) and with its interpretation
that this probability peak corresponds to a turning point where the
system is stopped prior to reaching the wall. At higher beam energies
the center of mass penetrates considerably through the wall (region
$X>0).$ As expected, this penetration is deeper for a more massive
non-interacting component; the peaks in the density inside the wall
can also be attributed to the non-interacting particle being stopped
via energy transfer to intrinsic excitations.

\section{Role of virtual channels and convergence\label{sec:Role-of-virtual}}

While problems similar to those presented here have been extensively
discussed in recent literature, for example Refs. \cite{Shegelski:2008i,Shegelski:2008,Razavy:2003,Kayanuma:1994,Hnybida:2008,Goodvin:2005,Bacca:2006},
little attention has been paid to virtual channels. In fact, most
of these works discuss tunneling of a diatomic molecule where both
the atoms interact with the potential. In that case, virtual excitations
are relatively less likely to take place, and hence the folded potential
within open channels already provides a relatively good description
of the process. Our selection of models on the other hand, where only
one particle interacts with the scatterer, is dynamically different.
It is the virtual channels that shape the non-interacting particle's
movement. Therefore, compared to the models discussed by other authors
cited above, our models are in general more sensitive to virtual channels.
In the most extreme case of reflection from an infinite wall, no meaningful
description is possible at all without reference to the virtual channels.
The folded potential for the ground state, depicted in Fig. \ref{ddynamic}(c)
with the solid black line, has a single hump, and therefore does not
lead to any resonant behavior in reactions at low energies, when only
one channel is open. Hence it can be concluded that, the resonance-like
increases or decreases in the transmission and reflection probabilities
shown, for example, in Figs. \ref{fig:VRA-HO_T} and \ref{fig:VRA-HO_R},
at low energies and especially when the non-interaction particle is
heavy, are exclusively due to virtual channels.

A successful extension of the VPM so as to include virtual channels
in the formalism, and the study of their role that we discuss in this
section, are among the main achievements of this work.

\begin{figure}
\includegraphics[width=7cm]{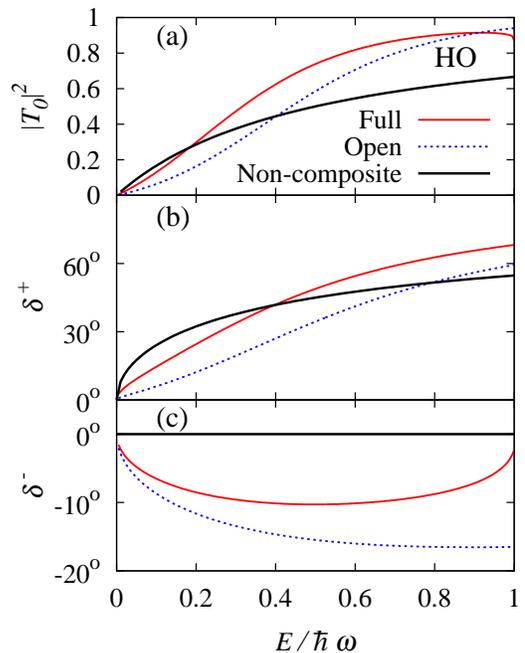}\caption{(Color online) \textbf{HO and $\delta$-barrier. }A two-body system
with $m_{1}=m_{2}$, bound by an oscillator confinement, is scattered
by an external delta potential with $E_{\delta}=1$. Panels (a), (b),
and (c) correspond to transmission probability, and two symmetric
and antisymmetric phase shifts, respectively. Three curves shown compare
full solution (labeled {}``full'') with the approximate treatment
that includes only the open channels ({}``open''), and with results
for non-composite projectile ({}``non-composite''). We conclude
that ignoring the composite nature of the system or neglecting the
virtual channels results in neither the phase shifts nor the transmission
probability correctly. \label{fig:virtualchannels}}

\end{figure}

The importance both of compositeness and of virtual channels is illustrated
in Fig. \ref{fig:virtualchannels}, where we consider an oscillator
scattering from a $\delta$-barrier. The curves in three different
styles and colors correspond to results from three different calculations:
the solid red line represents the exact solution, i.e., the solution
of scattering of a composite projectile obtained through a calculation
that includes the virtual channels as well as the open channels; the
dotted blue line represents scattering of a composite projectile solved
through a folded potential but ignoring any virtual channel whatsoever;
the solid black line represents scattering of a non-composite projectile
of the same mass as that of the composite projectile. For the region
of energies shown in these graphs, there is only one open channel;
thus the asymptotic behavior of the wave function is fully determined
by the two phase shifts $\delta^{\pm}$ which are shown in panels
(b) and (c) as functions of incident beam kinetic energy. All three
curves are different, indicating that neither non-compositeness nor
treatment of open channels only can substitute for a full solution.
To emphasize this, we show in panel (a) transmission probability,
an observable quantity, as a function of energy; for most of the energy
region shown, the actual transmission probability appears to be higher
than that of an equally massive non-composite particle.

\begin{figure}
\includegraphics[width=7cm]{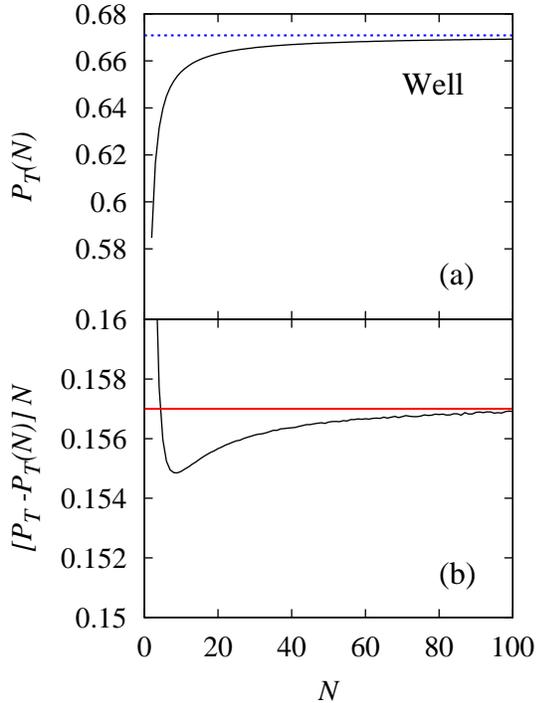}\caption{(Color online) \textbf{Well and $\delta$-barrier. }Convergence of
the transmission probability $P_{T}=|T_{0}|^{2}$ to its limiting
value, as a function of $N$. The masses are equal, $m_{1}=m{}_{2}$,
and the kinetic energy is $E=4$, so that only the first two channels
are open. The strength of the $\delta$-barrier is $E_{\delta}=1$.
Plot (a) shows the actual behavior of the transmission probability
$P_{T}(N)$ with $N$, and its asymptotic value $P_{T}=0.6708$ shown
by the grid-line. In order to show agreement with Eq. (\ref{eq:convergence}),
the value $N\left[P_{T}-P_{T}(N)\right]$ is shown in plot (b). This
quantity is well described by a constant $N_{s}=0.157,$ shown with
the red horizontal line. \label{fig:Convergence-of-transmission}}

\end{figure}

\begin{figure}
\includegraphics[width=7cm]{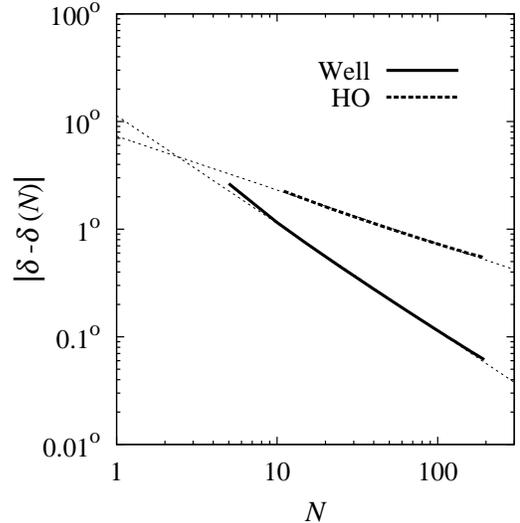}\caption{\textbf{Well/HO and wall. }Convergence of the phase shift. The log-log
scale acts to accentuate a good agreement with the $\sim1/|K_{N}|$
convergence rule. The dashed straight lines represent these rules.
For the square well intrinsic potential, $K_{N}\sim N$ and $\delta(N)\simeq\delta-N_{s}/N$
with $\delta=-23.05^{\circ}$ and $N_{s}=11.3.$ For the harmonic
oscillator $K_{N}\sim\sqrt{N}$, so $\delta(N)\simeq\delta-\sqrt{N_{s}/N}$
with $\delta=-22.98^{\circ}$ and $N_{s}=53.4$. For both cases the
incident beam energy is halfway between those of the ground state
and the first excited state.  \label{fig:Convergencerate}}

\end{figure}

While the virtual channels cannot in general be ignored, the contribution
of the highly excited states is expected to diminish. Practical applications
require some truncation in the channel space, too. In Fig. \ref{fig:Convergence-of-transmission}(a)
we demonstrate the rate of convergence of the transmission probability
$P_{T}=|T_{0}|^{2}$ by plotting it as a function of the number of
included channels $N$. The curve is visually indistinguishable from
the hyperbola \begin{equation}
P_{T}(N)=P_{T}-\frac{N_{s}}{N},\label{eq:convergence}\end{equation}
showing that the deviation of the probability $P_{T}(N)$
from its limiting value $P_{T}$ is inversely proportional to $N$.
That is, $N_{s}$ is the rate of convergence. The lower plot, Fig.
\ref{fig:Convergence-of-transmission}(b), shows agreement with Eq.
(\ref{eq:convergence}) by comparing $\left[P_{T}-P_{T}(N)\right]N$
with a constant $N_{s}$. These results are for the square well bound
system with $|K_{N}|\sim N$. With more precise consideration it is
found that in general, the amplitudes converge as $\sim1/|K_{N}|.$
Figure \ref{fig:Convergencerate} demonstrates an excellent agreement
with this rule using two different systems reflecting from an infinite
wall.

This power-law convergence is slow in contrast to an exponential convergence
usually encountered for eigenvalues and other structural observables
as functions of truncation \cite{Horoi:1999}. Here we repeat our
recent conjecture \cite{Ahsan:2010} that this is an inherent property
of reaction physics, where the kinetic energy operator plays a major
role in the Hamiltonian. The mentioned operator discretized in coordinate
space corresponds to a tri-diagonal matrix that meets a set of criteria
for the power-law convergence \cite{Horoi:1999}. 

In the course of our work we have vigorously tested the $\sim1/|K_{N}|$
convergence rule. While the rate of convergence, $N_{s}$, depends
strongly on the type of the system, we found no exception from the
power-law convergence.

\section{Intrinsic potential with a continuum; breakup\label{sec:continuum}}

We would like to conclude our exploration with a somewhat more realistic
situation where the intrinsic potential allows for a breakup. We therefore
consider a confining potential $v(x)$ that has both bound state(s)
and a continuum. In our discussion below we study a particular confinement,
namely, a finite square well (\textquotedbl{}finite well\textquotedbl{}):
\begin{equation}
v(x)=\left\{ \begin{array}{cc}
0 & \text{when}\,\,|x|>\lambda\\
-v & \text{otherwise}\end{array}\right.\,.\label{eq:FiniteSQWell}\end{equation}
This allows one to capture the generic features of the problem, while
still having a small number of parameters and retaining the ability
to have analytic solutions (\ref{eq:inHsolution}) for the intrinsic
Hamiltonian. Realistic applications to three-dimensional problems
with other potentials are outside the scope of this work.

In what follows we again select the units of length $\lambda$ to
represent the width of the intrinsic potential as defined in Eq. (\ref{eq:FiniteSQWell}).
The bound state energies for the finite square well potential are
given by the transcendental equation \begin{equation}
\tan\sqrt{\varepsilon+v}=\pm\left(\frac{-\varepsilon}{v+\varepsilon}\right)^{\pm1/2},\label{eq:ExactFSWenergy}\end{equation}
where we remind the reader that $v$ and $\varepsilon$ are expressed
in units of $\epsilon$ {[}see (\ref{eq:energyunit}){]}. The $\pm$
sign corresponds to the intrinsic parity ${\cal P}=\pm1$ of the state
of interest. In addition to these bound states there is a continuum
of states with positive energies above the well.

\begin{figure}
\includegraphics[width=7cm]{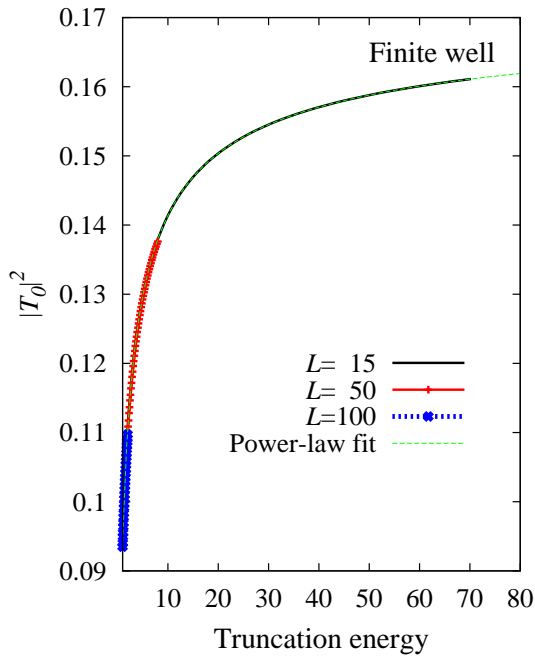}\caption{(Color online) \textbf{Finite well and $\delta$-barrier. }Convergence
of the transmission probability to its final value with truncation
energy (in units of $\epsilon$) of virtual excitations. The well-depth
is $v=1$ and the incident kinetic energy is $E=0.2$. Therefore,
only the elastic channel is open. The figure contains four different
curves, three of which are labeled with the widths of the quantization
box, $L=15,\,50,\:{\rm \text{{and}}}\:100$ that were used. These
curves agree well with the power-law convergence discussed in the
previous section, the fit for which is shown with the last curve (green,
dashed). The final value for transmission probability is 0.174. \label{fig:Contc}}

\end{figure}

To model such a situation mathematically we discretize the spectrum
using a quantization-box of width $2L$, and hence the intrinsic wave
functions $\psi_{n}(x)$ are subject to a boundary condition $\psi_{n}(\pm L)=0$.
The choice of a large enough $L$ can yield a spectrum that represents
the continuum obtained without the box. The large box allows for any
finite potentials $v(x)$ to be considered.

In order to examine the appropriateness of the approach and to address
the potential concerns arising from the presence of the continuum
and its truncation, we show in Fig. \ref{fig:Contc} the calculated
transmission probability for such a system incident on a $\delta$-barrier.
A depth of $v=1$ has been chosen for this example, thus there being
a single bound state at energy $\varepsilon_{0}=-0.454$, as follows
from Eq. (\ref{eq:ExactFSWenergy}), with the RMS size $1.17$ of
the wave function. The incident kinetic energy of the center-of-mass
motion is assumed to be $E=0.2$, which means that only the elastic
channel is open. Though there is not enough beam energy for a breakup,
the virtual channels are still important. In Fig. \ref{fig:Contc}
we explore different box-widths, $L=15,\,50,$ and $100$ (in units
of $\lambda$), which shows that the results are independent of $L$,
if it is large enough. (Note that with a large box-width the density
of states in the continuum is high and therefore it is difficult to
include high-energy channels.) Even for the smallest box $L=15$ the
energy of the ground state differs from the exact answer only by less
than 0.05\%. The breakup threshold is at $0.495$, which is slightly
different from $-\varepsilon_{0}$ mainly because the first excited
state (continuum threshold) in the box does not exactly coincide with
zero energy. These differences are minor and orders of magnitude smaller
for $L=100.$

\begin{figure}
\includegraphics[width=7cm]{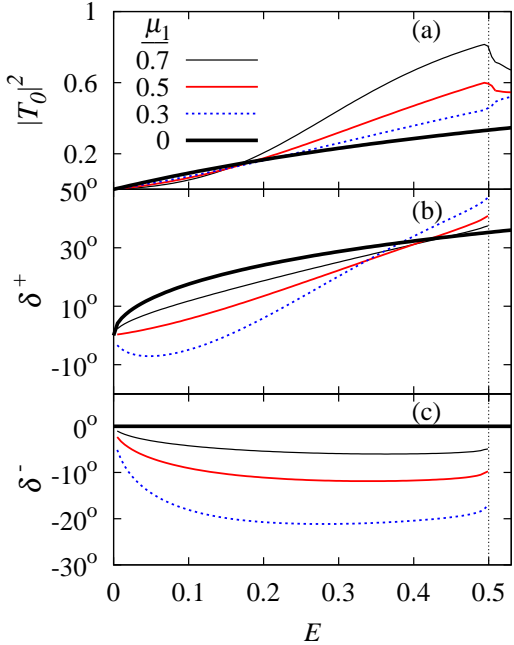}

\caption{(Color online) \textbf{Finite well and $\delta$-barrier. }Transmission
probability and phase shifts for the scattering of a two-particle
system with intrinsic potential that is finite and allows for breakup.
The potential is modeled by a square well of unit width and depth
$v=1.$ We concentrate on the region below the breakup threshold of
kinetic energy at $0.495$ shown with the vertical grid-line. Four
curves correspond to four different mass-ratios $\mu_{1}=0.3,\;0.5,\;\mathrm{and}\;0.7$
and the non composite limit of $0$ for which Eq. (\ref{eq:deltaamplitude})
is plotted. The $\delta$-barrier strength is assumed to be $E_{\delta}=1$.
The upper panel shows transmission probability; the curve is continued
above the breakup threshold to show the cusp at the threshold. The
two lower panels, similar to those in Figs. \ref{fig:VRA-HO_phase}
and \ref{fig:VRA-SW_phase}, show the phase shifts $\delta^{+}$ and
$\delta^{-}$ as defined in Eq. \ref{eq:define_phases}. Note that
their meanings as phase shifts of the S-matrix is true only below
the breakup threshold; above the breakup these are arguments of the
corresponding reflection and transmission amplitudes. \label{fig:contlow}}

\end{figure}

Now that the appropriateness and validity of the approach is established,
we present the transmission probability and both symmetric and antisymmetric
phase shifts $\delta^{\pm}$ as functions of incident kinetic energy
in Fig. \ref{fig:contlow}. This figure concentrates on the energy
region below the breakup threshold. This situation is important, since
it is commonly encountered in practice. As it is clear from the graphs,
the composite nature of the system and the virtual continuum are playing
a crucial role in shaping the reaction process. We find that at very
low energies transmission is inhibited for a composite particle. At
higher energies close to the breakup threshold, transmission rate
is always enhanced. Moreover, this rate increases for an increasingly
heavy non-interacting particle. The role of the virtual channels appears
to be universal for all models that we investigated {[}see also Figs.
\ref{fig:VRA-SW_T} and \ref{fig:VRA-HO_T}{]}, where $|T_{0}|^{2}$
increases sharply near the first threshold. The experiment in GANIL,
as cited at the end of Sec. \ref{sub:Results}, proves tunneling enhancement
for systems with breakup also.

\begin{figure}
\includegraphics[width=7cm]{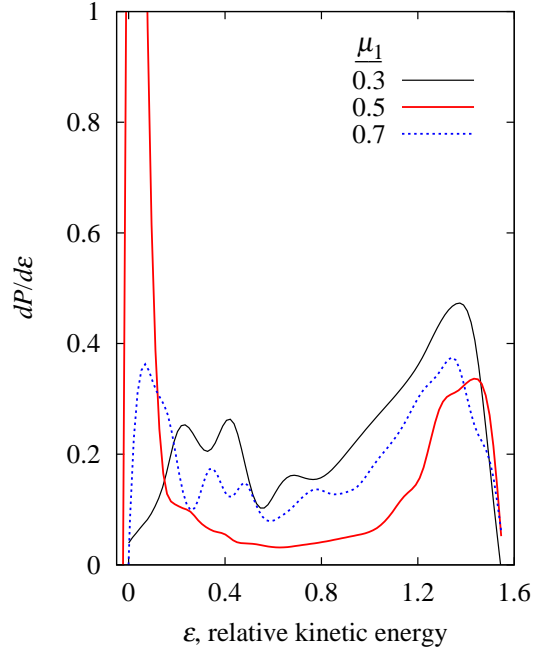}

\caption{(Color online) \textbf{Finite well and $\delta$-barrier. }Probability
to break up into a level $\varepsilon$ in the continuum is plotted
as a function of $\varepsilon$. The intrinsic potential is modeled
by a square well of unit width and depth $v=1$, as described in the
text. The kinetic energy of the incident wave is $E=2$. The system
interacts with a $\delta$-barrier of strength $E_{\delta}=1$, and
then either scatters elastically, or breaks up into its constituents
with relative energy $\varepsilon$. The three curves correspond to
three different mass-ratios $\mu_{1}=0.3,\;0.5,\;\mathrm{and}\;0.7$.
The corresponding probabilities for elastic scattering are $0.64$,
$0.60$, and $0.69$ respectively, which complement the breakup probabilities
shown here. \label{fig:cont_prob_dist_E}}

\end{figure}

It is interesting to review the distribution of probability of breakup
into the continuum, when energetically possible. In the continuum
we can still separate the center-of-mass and the relative kinetic
energies. The relative kinetic energy is now given by the discretized
states in the box. Fig. \ref{fig:cont_prob_dist_E} shows the normalized
probability distribution for breakup with different relative energies.
The initial beam in this case corresponds to a projectile in the ground
state with kinetic energy of $E=2$. Thus, the total kinetic energy
of fragments after the breakup is ${E}+\varepsilon_{0}-\varepsilon_{1}\approx{E}+\varepsilon_{0}=1.55$
which is viewed as a sum of two parts: the center-of-mass kinetic
energy $\hbar^{2}K^{2}/(2M)$ and the relative kinetic energy $\varepsilon$.
As seen from the plot, it is most likely to have the two fragments
moving together with very little relative energy (corresponding to
the peaks on the left side of the plot), or, inversely, moving apart
in opposite directions with most of the energy concentrated in the
relative motion (corresponding to the peaks on the right side).

\begin{figure}
\includegraphics[width=7cm]{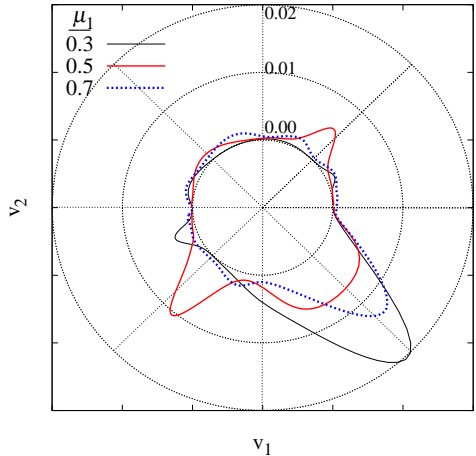}

\caption{(Color online) \textbf{Finite well and $\delta$-barrier. }The system
and scattering conditions are the same as in Fig. \ref{fig:cont_prob_dist_E}.
The system interacts with the $\delta$-barrier, and then either scatters
elastically, or breaks up into its constituents with momenta $\mathrm{v}{}_{1}$
and $\mathrm{v}_{2}$. Break-up probability, as a function of velocities
$\mathrm{v}_{1}$ and $\mathrm{v}{}_{2}$, is plotted as a contour
graph. Due to Eq. (\ref{eq:ellipse}) the points representing the
possible velocity values lie on an ellipse in the $\mathrm{(v}{}_{1},\,\mathrm{v}{}_{2})$
plane and are parametrized by a single angle. The contour plot shows
the probability of breakup (radial coordinate) as a function of this
angle. \label{fig:cont_contour}}

\end{figure}

The choice of quantities plotted in Fig. \ref{fig:cont_prob_dist_E}
reflect our method, but is not very closely related to a potential
experiment, where the momenta or velocities of both particles could
be measured. Therefore in Fig. \ref{fig:cont_contour} we show the
distribution of probability to observe a certain combination of the
particle-velocities $\mathrm{v}_{1}$ and $\mathrm{v}_{2}$. Since
the total kinetic energy after breakup is fixed, \begin{equation}
\frac{1}{2}\left(m_{1}\mathrm{v}_{1}^{2}+m_{2}\mathrm{v}_{2}^{2}\right)=E+\varepsilon_{0},\label{eq:ellipse}\end{equation}
it is sufficient to use a single angle to parametrize the position
on the ellipse formed in the velocity $(\mathrm{v}_{1},\,\mathrm{v}_{2})$
plane. In Fig. \ref{fig:cont_contour} we show the probability as
a function of angle using a contour plot. Both Figures \ref{fig:cont_prob_dist_E}
and \ref{fig:cont_contour} have been smoothed, but preserve the general
shapes which we believe to be good representations of the physics.
Some of the features seen in Fig. \ref{fig:cont_prob_dist_E} become
more transparent in Fig. \ref{fig:cont_contour}. For all mass-ratios
we see that the probability peaks at about $-45^{\circ}$ which corresponds
to non-interacting particle-1 moving forward and interacting particle-2
being reflected back with velocities nearly equal in magnitude and
opposite in direction. For equal masses the peak at low relative kinetic
energies in Fig. \ref{fig:cont_prob_dist_E} appears as two peaks
in Fig. \ref{fig:cont_contour} at about $45^{\circ}$ and $-135^{\circ}$.
In both cases the particles move with similar velocities, forward
for $45^{\circ}$ and backward for $-135^{\circ}.$ The observed picture
appears to be quite intuitive.

\section{Summary and conclusions \label{sec:Summary-and-conclusions}}

In this work we revisit some of the most intricate questions of reaction
physics involving composite objects. The research presented here was
inspired by highly unpredictable behavior of reaction observables
including resonances and cusp-discontinuities, a very broad spectrum
of scales involved, and, at the same time, the utmost importance of
these processes in nuclear physics and other fields. 

The topic of reactions involving composite objects is widely investigated,
and benefits from many advanced methods and techniques. Our work got
its thrust from a simple and well-defined problem of a deuteron-like
system interacting in one dimension with an infinite Coulomb wall.
A number of methods tried, in the past, for this problem have failed
under certain circumstances, namely, where approximations or simplifications
commonly used were not appropriate for this particular problem, or
where the method did not produce convergent results, or where there
were difficulties with numerical errors. Due to this delicate nature
of the problem, we find \emph{exact solutions} to all the examples
considered in this work. This allows us to obtain comprehensive answers,
and to be able to carry out comparisons with other methods and solutions.

This goal requires us to have a precisely-defined Hamiltonian with
as few parameters as possible, and conditions that perhaps are more
critical to reaction-structure interplay than those typically encountered
in nature. In this presentation we restrict our discussion to models.
Nevertheless, our methods have broad applicability; most examples
can be modified easily to represent realistic situations, and we continuously
suggest cases in nature that are similar to what we discuss. We study,
through this work, a two-particle system interacting in a one-dimensional
scattering with a target that poses a $\delta$-potential or an infinite
wall potential. It is always assumed that only one of the two components
interacts with the target. The study includes models that do allow
the projectile to breakup, and models that do not. The dominant and
non-perturbative role of the virtual channels that extend far in excitation
energy is the main common theme of all the examples discussed here.

We start by revisiting the \textquotedbl{}deuteron and Coulomb-wall\textquotedbl{}
model which has been discussed for almost a decade with little outcome
\cite{MORO:2000,Sakharuk:2000,Ahsan:2007}. Unfortunately, this problem
is commonly dismissed either at the first glance when it seems uninteresting,
or after some investigation when it seems unphysical, ill-defined,
or unsolvable. We, on the other hand, find this model remarkable in
its ability to demonstrate, in an extremely transparent manner, the
dynamics driven by the virtual excitations.

We review, and carefully apply, the technique of projecting the reaction
dynamics onto an intrinsic space and show that while satisfactory
results are obtained in some limits, this formally exact approach
does not yield convergent solutions in general. This is an important
finding because this ``Projection Method'' is a prototype of several
commonly used approaches in many-body problems that involve both structure
and reactions \cite{Okolowicz:2003,Volya:2006}. 

As our main workhorse we utilize the Variable Phase Method (VPM) to
address the coupled-channel problems of interest. While the method
has been used by others before, we modify and extend it to treat highly
remote virtual channels. We demonstrate that the VPM produces reliable
and convergent results. We investigate the contributions from remote
virtual excitations, study convergence with the truncation size, and
find the power-law convergence, which is in contrast to the exponential
convergence seen in many-body structure problems \cite{Horoi:1999}.

Within a given set of models, this work contains numerous examples,
investigations and demonstrations. Cusps and discontinuities appear
in observables as manifestations of conservation of probability and
redistribution of flux at the thresholds. Intrinsic structure gives
rise to resonance-like behavior in tunneling probabilities; our models
and recent experimental evidences indicate a generic enhancement in
transmission probabilities due to virtual channels or a virtual continuum,
whichever is the case. We explore and discuss the role of virtual
excitations at very low energies, showing that even in those cases
the scattering length is sensitive to the projectile's structure.
Due to the intrinsic structure and its coupling to reaction dynamics,
scattering length can become infinite, the phenomenon being known
as Feshbach resonance. We demonstrate how the intrinsic structure
violates charge symmetry, which is called the Barkas effect. The scattering
of a non-composite projectile off a $\delta$-barrier is the same
for attractive and repulsive interactions. But, in case of a composite
projectile, the corresponding three-body problem for an attractive
potential is quite different from that for a repulsive barrier, and
reveals numerous resonances, some of which can be understood as bound
states built upon individual intrinsic excitations involving two-body
subsystems.

Scattering and breakup dynamics influenced by a virtual continuum
are also investigated in this work. It is seen that the most probable
breakups take place where either almost all the kinetic energy is
relative, or almost all of it is in the center of mass.

\begin{acknowledgments}
We are thankful to C. Bertulani, M. Horoi, A. Moro, A. Sakharuk, and
V. Zelevinsky for bringing this topic to our attention and for years
of motivating discussions. We also acknowledge the support from the
U. S. Department of Energy under the DE-FG02-92ER40750 grant. 
\end{acknowledgments}

\bibliographystyle{apsrev}

\begin{thebibliography}{47}
\expandafter\ifx\csname natexlab\endcsname\relax\def\natexlab#1{#1}\fi
\expandafter\ifx\csname bibnamefont\endcsname\relax
  \def\bibnamefont#1{#1}\fi
\expandafter\ifx\csname bibfnamefont\endcsname\relax
  \def\bibfnamefont#1{#1}\fi
\expandafter\ifx\csname citenamefont\endcsname\relax
  \def\citenamefont#1{#1}\fi
\expandafter\ifx\csname url\endcsname\relax
  \def\url#1{\texttt{#1}}\fi
\expandafter\ifx\csname urlprefix\endcsname\relax\def\urlprefix{URL }\fi
\providecommand{\bibinfo}[2]{#2}
\providecommand{\eprint}[2][]{\url{#2}}

\bibitem[{\citenamefont{Flambaum and Zelevinsky}(2005)}]{Flambaum:2005}
\bibinfo{author}{\bibfnamefont{V.~V.} \bibnamefont{Flambaum}} \bibnamefont{and}
  \bibinfo{author}{\bibfnamefont{V.~G.} \bibnamefont{Zelevinsky}},
  \bibinfo{journal}{J. Phys. G: Nucl. Part. Phys.}
  \textbf{\bibinfo{volume}{31}}, \bibinfo{pages}{355} (\bibinfo{year}{2005}).

\bibitem[{\citenamefont{Bertulani et~al.}(2007)\citenamefont{Bertulani,
  Flambaum, and Zelevinsky}}]{Bertulani:2007}
\bibinfo{author}{\bibfnamefont{C.~A.} \bibnamefont{Bertulani}},
  \bibinfo{author}{\bibfnamefont{V.~V.} \bibnamefont{Flambaum}},
  \bibnamefont{and} \bibinfo{author}{\bibfnamefont{V.~G.}
  \bibnamefont{Zelevinsky}}, \bibinfo{journal}{J. Phys. G: Nucl. Part. Phys.}
  \textbf{\bibinfo{volume}{34}}, \bibinfo{pages}{2289} (\bibinfo{year}{2007}).

\bibitem[{\citenamefont{Balantekin and Takigawa}(1998)}]{Balantekin:1998}
\bibinfo{author}{\bibfnamefont{A.~B.} \bibnamefont{Balantekin}}
  \bibnamefont{and} \bibinfo{author}{\bibfnamefont{N.}~\bibnamefont{Takigawa}},
  \bibinfo{journal}{Rev. Mod. Phys.} \textbf{\bibinfo{volume}{70}},
  \bibinfo{pages}{77} (\bibinfo{year}{1998}).

\bibitem[{\citenamefont{Bonini et~al.}(1999)\citenamefont{Bonini, Cohen, Rebbi,
  and Rubakov}}]{Bonini:1999}
\bibinfo{author}{\bibfnamefont{G.~F.} \bibnamefont{Bonini}},
  \bibinfo{author}{\bibfnamefont{A.}~\bibnamefont{Cohen}},
  \bibinfo{author}{\bibfnamefont{C.}~\bibnamefont{Rebbi}}, \bibnamefont{and}
  \bibinfo{author}{\bibfnamefont{V.}~\bibnamefont{Rubakov}},
  \bibinfo{journal}{Phys. Rev. D} \textbf{\bibinfo{volume}{60}},
  \bibinfo{pages}{076004} (\bibinfo{year}{1999}).

\bibitem[{\citenamefont{Goodvin and Shegelski}(2005)}]{Goodvin:2005}
\bibinfo{author}{\bibfnamefont{G.~L.} \bibnamefont{Goodvin}} \bibnamefont{and}
  \bibinfo{author}{\bibfnamefont{M.~R.~A.} \bibnamefont{Shegelski}},
  \bibinfo{journal}{Phys. Rev. A} \textbf{\bibinfo{volume}{72}},
  \bibinfo{pages}{042713} (\bibinfo{year}{2005}).

\bibitem[{\citenamefont{Saito and Kayanuma}(1994)}]{Kayanuma:1994}
\bibinfo{author}{\bibfnamefont{N.}~\bibnamefont{Saito}} \bibnamefont{and}
  \bibinfo{author}{\bibfnamefont{Y.}~\bibnamefont{Kayanuma}},
  \bibinfo{journal}{J. Phys.: Condens. Matter} \textbf{\bibinfo{volume}{6}},
  \bibinfo{pages}{3759} (\bibinfo{year}{1994}).

\bibitem[{\citenamefont{Bacca and Feldmeier}(2006)}]{Bacca:2006}
\bibinfo{author}{\bibfnamefont{S.}~\bibnamefont{Bacca}} \bibnamefont{and}
  \bibinfo{author}{\bibfnamefont{H.}~\bibnamefont{Feldmeier}},
  \bibinfo{journal}{Phys. Rev. C} \textbf{\bibinfo{volume}{73}},
  \bibinfo{pages}{054608} (\bibinfo{year}{2006}).

\bibitem[{\citenamefont{Ivlev and Gudkov}(2004)}]{Ivlev:2004}
\bibinfo{author}{\bibfnamefont{B.}~\bibnamefont{Ivlev}} \bibnamefont{and}
  \bibinfo{author}{\bibfnamefont{V.}~\bibnamefont{Gudkov}},
  \bibinfo{journal}{Phys. Rev. C} \textbf{\bibinfo{volume}{69}},
  \bibinfo{pages}{037602} (\bibinfo{year}{2004}).

\bibitem[{\citenamefont{Flambaum and Zelevinsky}(1999)}]{Flambaum:1999}
\bibinfo{author}{\bibfnamefont{V.~V.} \bibnamefont{Flambaum}} \bibnamefont{and}
  \bibinfo{author}{\bibfnamefont{V.~G.} \bibnamefont{Zelevinsky}},
  \bibinfo{journal}{Phys. Rev. Lett.} \textbf{\bibinfo{volume}{83}},
  \bibinfo{pages}{3108} (\bibinfo{year}{1999}).

\bibitem[{\citenamefont{Ahsan and Volya}(2010)}]{Ahsan:2010}
\bibinfo{author}{\bibfnamefont{N.}~\bibnamefont{Ahsan}} \bibnamefont{and}
  \bibinfo{author}{\bibfnamefont{A.}~\bibnamefont{Volya}}, in
  \emph{\bibinfo{booktitle}{JPCS, proceedings of the International Nuclear
  Physics Conference 2010, Vancouver, Canada}} (\bibinfo{year}{2010}).

\bibitem[{\citenamefont{Lemasson et~al.}(2009)\citenamefont{Lemasson,
  Shrivastava, Navin, Rejmund, Keeley, Zelevinsky, Bhattacharyya, Chatterjee,
  de~France, Jacquot et~al.}}]{Lemasson:2009}
\bibinfo{author}{\bibfnamefont{A.}~\bibnamefont{Lemasson}},
  \bibinfo{author}{\bibfnamefont{A.}~\bibnamefont{Shrivastava}},
  \bibinfo{author}{\bibfnamefont{A.}~\bibnamefont{Navin}},
  \bibinfo{author}{\bibfnamefont{M.}~\bibnamefont{Rejmund}},
  \bibinfo{author}{\bibfnamefont{N.}~\bibnamefont{Keeley}},
  \bibinfo{author}{\bibfnamefont{V.}~\bibnamefont{Zelevinsky}},
  \bibinfo{author}{\bibfnamefont{S.}~\bibnamefont{Bhattacharyya}},
  \bibinfo{author}{\bibfnamefont{A.}~\bibnamefont{Chatterjee}},
  \bibinfo{author}{\bibfnamefont{G.}~\bibnamefont{de~France}},
  \bibinfo{author}{\bibfnamefont{B.}~\bibnamefont{Jacquot}},
  \bibnamefont{et~al.}, \bibinfo{journal}{Phys. Rev. Lett.}
  \textbf{\bibinfo{volume}{103}}, \bibinfo{pages}{232701} (\bibinfo{year}{2009}).

\bibitem[{\citenamefont{Merzbacher}(1998)}]{Merzbacher:1998}
\bibinfo{author}{\bibfnamefont{E.}~\bibnamefont{Merzbacher}},
  \emph{\bibinfo{title}{Quantum Mechanics}} (\bibinfo{publisher}{John Wiley and
  Sons, Inc.}, \bibinfo{year}{1998}).

\bibitem[{\citenamefont{Lipkin}(1973)}]{Lipkin:1973}
\bibinfo{author}{\bibfnamefont{H.~J.} \bibnamefont{Lipkin}},
  \emph{\bibinfo{title}{Quantum Mechanics: New Approaches to Selected Topics}}
  (\bibinfo{publisher}{North-Holland Pub. Co.}, \bibinfo{address}{Amsterdam},
  \bibinfo{year}{1973}).

\bibitem[{\citenamefont{Nogami and Ross}(1996)}]{Nogami:1996}
\bibinfo{author}{\bibfnamefont{Y.}~\bibnamefont{Nogami}} \bibnamefont{and}
  \bibinfo{author}{\bibfnamefont{C.~K.} \bibnamefont{Ross}},
  \bibinfo{journal}{Am. J. Phys.} \textbf{\bibinfo{volume}{64}},
  \bibinfo{pages}{923} (\bibinfo{year}{1996}).

\bibitem[{\citenamefont{Kiers and van Dijk}(1996)}]{Kiers:1996}
\bibinfo{author}{\bibfnamefont{K.~A.} \bibnamefont{Kiers}} \bibnamefont{and}
  \bibinfo{author}{\bibfnamefont{W.}~\bibnamefont{van Dijk}},
  \bibinfo{journal}{J. Math. Phys.} \textbf{\bibinfo{volume}{37}},
  \bibinfo{pages}{6033} (\bibinfo{year}{1996}).

\bibitem[{\citenamefont{van Dijk et~al.}(2008)\citenamefont{van Dijk, Spyksma,
  and West}}]{van_Dijk:2008}
\bibinfo{author}{\bibfnamefont{W.}~\bibnamefont{van Dijk}},
  \bibinfo{author}{\bibfnamefont{K.}~\bibnamefont{Spyksma}}, \bibnamefont{and}
  \bibinfo{author}{\bibfnamefont{M.}~\bibnamefont{West}},
  \bibinfo{journal}{Phys. Rev. A} \textbf{\bibinfo{volume}{78}},
  \bibinfo{pages}{022108} (\bibinfo{year}{2008}).

\bibitem[{\citenamefont{Baz et~al.}(1971)\citenamefont{Baz, Zeldovich, and
  Perelomov}}]{Baz:1971}
\bibinfo{author}{\bibfnamefont{A.~I.} \bibnamefont{Baz}},
  \bibinfo{author}{\bibfnamefont{Y.~B.} \bibnamefont{Zeldovich}},
  \bibnamefont{and} \bibinfo{author}{\bibfnamefont{A.~M.}
  \bibnamefont{Perelomov}}, \emph{\bibinfo{title}{Scattering, Reactions and
  Decays in Nonrelativistic Quantum Mechanics}} (\bibinfo{publisher}{Nauka,
  Moscow}, \bibinfo{year}{1971}).

\bibitem[{\citenamefont{Moro et~al.}(2000)\citenamefont{Moro, Caballero, and
  G\'omez-Camacho}}]{MORO:2000}
\bibinfo{author}{\bibfnamefont{A.~M.} \bibnamefont{Moro}},
  \bibinfo{author}{\bibfnamefont{J.~A.} \bibnamefont{Caballero}},
  \bibnamefont{and} \bibinfo{author}{\bibnamefont{G\'omez-Camacho}},
  \emph{\bibinfo{title}{One dimensional scattering of a two body interacting
  system by an infinite wall}} (\bibinfo{year}{2000}),
  \bibinfo{note}{unpublished}.

\bibitem[{\citenamefont{Sakharuk and Zelevinsky}(1999)}]{Sakharuk:1999}
\bibinfo{author}{\bibfnamefont{A.}~\bibnamefont{Sakharuk}} \bibnamefont{and}
  \bibinfo{author}{\bibfnamefont{V.}~\bibnamefont{Zelevinsky}}, in
  \emph{\bibinfo{booktitle}{APS Ohio Section Fall Meeting}}
  (\bibinfo{year}{1999}), \bibinfo{note}{1999APS..OSF..CD09S}.

\bibitem[{\citenamefont{Ahsan and Volya}(2007)}]{Ahsan:2007}
\bibinfo{author}{\bibfnamefont{N.}~\bibnamefont{Ahsan}} \bibnamefont{and}
  \bibinfo{author}{\bibfnamefont{A.}~\bibnamefont{Volya}}, in
  \emph{\bibinfo{booktitle}{Changing Facets of Nuclear Structure, proceedings
  of the 9th International Spring Seminar on Nuclear Physics, Vico Equense,
  Italy, May 2007}}, edited by
  \bibinfo{editor}{\bibfnamefont{A.}~\bibnamefont{Covello}}
  (\bibinfo{publisher}{World Scientific Publishing Co. Pte. Ltd.},
  \bibinfo{year}{2007}).

\bibitem[{\citenamefont{Okolowicz et~al.}(2003)\citenamefont{Okolowicz,
  Ploszajczak, and Rotter}}]{Okolowicz:2003}
\bibinfo{author}{\bibfnamefont{J.}~\bibnamefont{Okolowicz}},
  \bibinfo{author}{\bibfnamefont{M.}~\bibnamefont{Ploszajczak}},
  \bibnamefont{and} \bibinfo{author}{\bibfnamefont{I.}~\bibnamefont{Rotter}},
  \bibinfo{journal}{Phys. Rep.} \textbf{\bibinfo{volume}{374}},
  \bibinfo{pages}{271} (\bibinfo{year}{2003}).

\bibitem[{\citenamefont{Volya and Zelevinsky}(2006)}]{Volya:2006}
\bibinfo{author}{\bibfnamefont{A.}~\bibnamefont{Volya}} \bibnamefont{and}
  \bibinfo{author}{\bibfnamefont{V.}~\bibnamefont{Zelevinsky}},
  \bibinfo{journal}{Phys. Rev. C} \textbf{\bibinfo{volume}{74}},
  \bibinfo{pages}{064314} (\bibinfo{year}{2006}).

\bibitem[{\citenamefont{Volya}(2009)}]{Volya:2009}
\bibinfo{author}{\bibfnamefont{A.}~\bibnamefont{Volya}},
  \bibinfo{journal}{Phys. Rev. C} \textbf{\bibinfo{volume}{79}},
  \bibinfo{pages}{044308} (\bibinfo{year}{2009}).

\bibitem[{\citenamefont{Horoi et~al.}(1999)\citenamefont{Horoi, Volya, and
  Zelevinsky}}]{Horoi:1999}
\bibinfo{author}{\bibfnamefont{M.}~\bibnamefont{Horoi}},
  \bibinfo{author}{\bibfnamefont{A.}~\bibnamefont{Volya}}, \bibnamefont{and}
  \bibinfo{author}{\bibfnamefont{V.}~\bibnamefont{Zelevinsky}},
  \bibinfo{journal}{Phys. Rev. Lett.} \textbf{\bibinfo{volume}{82}},
  \bibinfo{pages}{2064} (\bibinfo{year}{1999}).

\bibitem[{\citenamefont{Sakharuk}()}]{Sakharuk:2000}
\bibinfo{author}{\bibfnamefont{A.}~\bibnamefont{Sakharuk}},
  \bibinfo{note}{private communication}.

\bibitem[{\citenamefont{Zelevinsky}()}]{Zelevinsky:2000}
\bibinfo{author}{\bibfnamefont{V.}~\bibnamefont{Zelevinsky}},
  \bibinfo{note}{private communication}.

\bibitem[{\citenamefont{Zelevinsky and Sakharuk}(2005)}]{Zelevinsky:2005}
\bibinfo{author}{\bibfnamefont{V.}~\bibnamefont{Zelevinsky}} \bibnamefont{and}
  \bibinfo{author}{\bibfnamefont{A.}~\bibnamefont{Sakharuk}}, in
  \emph{\bibinfo{booktitle}{Bulletin of the American Physical Society. 2005 APS
  April Meeting}} (\bibinfo{year}{2005}), \bibinfo{note}{bAPS.2005.APR.C13.2}.

\bibitem[{\citenamefont{Horoi}()}]{Horoi:2000}
\bibinfo{author}{\bibfnamefont{M.}~\bibnamefont{Horoi}}, \bibinfo{note}{private
  communication}.

\bibitem[{\citenamefont{Landau and Lifshitz}(1981)}]{Landau:1981}
\bibinfo{author}{\bibfnamefont{L.~D.} \bibnamefont{Landau}} \bibnamefont{and}
  \bibinfo{author}{\bibfnamefont{E.~M.} \bibnamefont{Lifshitz}},
  \emph{\bibinfo{title}{Quantum Mechanics. Non-relativistic theory.}}
  (\bibinfo{publisher}{Pergamon Press, New York}, \bibinfo{year}{1981}).

\bibitem[{\citenamefont{Razavy}(2003)}]{Razavy:2003}
\bibinfo{author}{\bibfnamefont{M.}~\bibnamefont{Razavy}},
  \emph{\bibinfo{title}{Quantum Theory of Tunneling}}
  (\bibinfo{publisher}{World Scientific Publishing Co. Pte. Ltd.},
  \bibinfo{year}{2003}).

\bibitem[{\citenamefont{Morse and Allis}(1933)}]{Morse:1933}
\bibinfo{author}{\bibfnamefont{P.~M.} \bibnamefont{Morse}} \bibnamefont{and}
  \bibinfo{author}{\bibfnamefont{W.~P.} \bibnamefont{Allis}},
  \bibinfo{journal}{Phys. Rev.} \textbf{\bibinfo{volume}{44}},
  \bibinfo{pages}{269} (\bibinfo{year}{1933}).

\bibitem[{\citenamefont{Drukarev}(1949)}]{Drukarev:1949}
\bibinfo{author}{\bibfnamefont{G.~F.} \bibnamefont{Drukarev}},
  \bibinfo{journal}{Zh. Eksp. Teor. Fiz.} \textbf{\bibinfo{volume}{19}},
  \bibinfo{pages}{247} (\bibinfo{year}{1949}).

\bibitem[{\citenamefont{Kynch}(1952)}]{Kynch:1952}
\bibinfo{author}{\bibfnamefont{G.~J.} \bibnamefont{Kynch}}, in
  \emph{\bibinfo{booktitle}{Proc. Phys. Soc. A}} (\bibinfo{year}{1952}),
  vol.~\bibinfo{volume}{65}, p. \bibinfo{pages}{708}.

\bibitem[{\citenamefont{Calogero}(1963)}]{Calogero:1963}
\bibinfo{author}{\bibfnamefont{F.}~\bibnamefont{Calogero}},
  \bibinfo{journal}{Nuovo Cimento} \textbf{\bibinfo{volume}{27}},
  \bibinfo{pages}{947} (\bibinfo{year}{1963}).

\bibitem[{\citenamefont{Calogero and Ravenhall}(1964)}]{Calogero:1964}
\bibinfo{author}{\bibfnamefont{F.}~\bibnamefont{Calogero}} \bibnamefont{and}
  \bibinfo{author}{\bibfnamefont{D.~G.} \bibnamefont{Ravenhall}},
  \bibinfo{journal}{Nuovo Cimento} \textbf{\bibinfo{volume}{32}},
  \bibinfo{pages}{1755} (\bibinfo{year}{1964}).

\bibitem[{\citenamefont{Babikov}(1967)}]{Babikov:1967}
\bibinfo{author}{\bibfnamefont{V.~V.} \bibnamefont{Babikov}},
  \bibinfo{journal}{Sov. Phys.-Usp.} \textbf{\bibinfo{volume}{10}},
  \bibinfo{pages}{271} (\bibinfo{year}{1967}).

\bibitem[{\citenamefont{Tikochinsky}(1970)}]{Tikochinsky:1970}
\bibinfo{author}{\bibfnamefont{Y.}~\bibnamefont{Tikochinsky}},
  \bibinfo{journal}{J. Math. Phys.} \textbf{\bibinfo{volume}{11}},
  \bibinfo{pages}{3019} (\bibinfo{year}{1970}).

\bibitem[{\citenamefont{Babikov}(1968)}]{Babikov:1968}
\bibinfo{author}{\bibfnamefont{V.~V.} \bibnamefont{Babikov}},
  \emph{\bibinfo{title}{Method Fazovych Funkzii v Kvantovoi Mechanike (Method
  of Phase Functions in Quantum Mechanics)}} (\bibinfo{publisher}{Nauka},
  \bibinfo{address}{Moscow}, \bibinfo{year}{1968}).

\bibitem[{\citenamefont{Calogero}(1967)}]{Calogero:1967}
\bibinfo{author}{\bibfnamefont{F.}~\bibnamefont{Calogero}},
  \emph{\bibinfo{title}{Variable Phase Approach to Potential Scattering}},
  vol.~\bibinfo{volume}{35} (\bibinfo{publisher}{Academic Press},
  \bibinfo{address}{New York}, \bibinfo{year}{1967}).

\bibitem[{\citenamefont{Hnybida and Shegelski}(2008)}]{Hnybida:2008}
\bibinfo{author}{\bibfnamefont{J.}~\bibnamefont{Hnybida}} \bibnamefont{and}
  \bibinfo{author}{\bibfnamefont{M.~R.~A.} \bibnamefont{Shegelski}},
  \bibinfo{journal}{Phys. Rev. A} \textbf{\bibinfo{volume}{78}},
  \bibinfo{pages}{032711} (\bibinfo{year}{2008}).

\bibitem[{\citenamefont{Shegelski
  et~al.}(2008{\natexlab{a}})\citenamefont{Shegelski, Hnybida, Friesen, Lind,
  and Kavka}}]{Shegelski:2008}
\bibinfo{author}{\bibfnamefont{M.~R.~A.} \bibnamefont{Shegelski}},
  \bibinfo{author}{\bibfnamefont{J.}~\bibnamefont{Hnybida}},
  \bibinfo{author}{\bibfnamefont{H.}~\bibnamefont{Friesen}},
  \bibinfo{author}{\bibfnamefont{C.}~\bibnamefont{Lind}}, \bibnamefont{and}
  \bibinfo{author}{\bibfnamefont{J.}~\bibnamefont{Kavka}},
  \bibinfo{journal}{Phys. Rev. A} \textbf{\bibinfo{volume}{77}},
  \bibinfo{pages}{032702} (\bibinfo{year}{2008}{\natexlab{a}}).

\bibitem[{\citenamefont{Shegelski
  et~al.}(2008{\natexlab{b}})\citenamefont{Shegelski, Hnybida, and
  Vogt}}]{Shegelski:2008i}
\bibinfo{author}{\bibfnamefont{M.~R.~A.} \bibnamefont{Shegelski}},
  \bibinfo{author}{\bibfnamefont{J.}~\bibnamefont{Hnybida}}, \bibnamefont{and}
  \bibinfo{author}{\bibfnamefont{R.}~\bibnamefont{Vogt}},
  \bibinfo{journal}{Phys. Rev. A} \textbf{\bibinfo{volume}{78}},
  \bibinfo{pages}{062703} (\bibinfo{year}{2008}{\natexlab{b}}).

\bibitem[{\citenamefont{Talukdar et~al.}(1981)\citenamefont{Talukdar, Mallick,
  and Roy}}]{Talukdar:1981}
\bibinfo{author}{\bibfnamefont{B.}~\bibnamefont{Talukdar}},
  \bibinfo{author}{\bibfnamefont{N.}~\bibnamefont{Mallick}}, \bibnamefont{and}
  \bibinfo{author}{\bibfnamefont{D.}~\bibnamefont{Roy}}, \bibinfo{journal}{J.
  Phys. G: Nucl. Part. Phys.} \textbf{\bibinfo{volume}{7}},
  \bibinfo{pages}{1103} (\bibinfo{year}{1981}).

\bibitem[{\citenamefont{Tikochinsky}(1977)}]{Tikochinsky:1977}
\bibinfo{author}{\bibfnamefont{Y.}~\bibnamefont{Tikochinsky}},
  \bibinfo{journal}{Am. J. Phys.} \textbf{\bibinfo{volume}{103}},
  \bibinfo{pages}{185} (\bibinfo{year}{1977}).

\bibitem[{\citenamefont{Volya and Esbensen}(2002)}]{Volya:2002}
\bibinfo{author}{\bibfnamefont{A.}~\bibnamefont{Volya}} \bibnamefont{and}
  \bibinfo{author}{\bibfnamefont{H.}~\bibnamefont{Esbensen}},
  \bibinfo{journal}{Phys. Rev. C} \textbf{\bibinfo{volume}{66}},
  \bibinfo{pages}{044604} (\bibinfo{year}{2002}).

\bibitem[{\citenamefont{Levy and Keller}(1963)}]{Levy:1963}
\bibinfo{author}{\bibfnamefont{B.~R.} \bibnamefont{Levy}} \bibnamefont{and}
  \bibinfo{author}{\bibfnamefont{J.~B.} \bibnamefont{Keller}},
  \bibinfo{journal}{J. Math. Phys.} \textbf{\bibinfo{volume}{4}},
  \bibinfo{pages}{54} (\bibinfo{year}{1963}).

\bibitem[{\citenamefont{Dashen}(1963)}]{Dashen:1963}
\bibinfo{author}{\bibfnamefont{R.~F.} \bibnamefont{Dashen}},
  \bibinfo{journal}{J. Math. Phys.} \textbf{\bibinfo{volume}{4}},
  \bibinfo{pages}{388} (\bibinfo{year}{1963}).

\end{thebibliography}

\end{document}